\documentstyle[12pt,aaspp4]{article}

\slugcomment{}

\lefthead{S.E. Church et al.}
\righthead{CMB anisotropies on arcminute scales}

\begin{document}

\title{An Upper Limit to Arcminute Scale Anisotropy in the Cosmic
  Microwave Background Radiation at 142\,GHz}

\author{
  S.E.~Church\altaffilmark{1},
  K.M.~Ganga\altaffilmark{1,2},
  P.A.R.~Ade\altaffilmark{3},
  W.L.~Holzapfel\altaffilmark{1,4,5},
  P.D.~Mauskopf\altaffilmark{1,4,6},
  T.M.~Wilbanks\altaffilmark{4,7} \&
  A.E.~Lange\altaffilmark{1}
}

\altaffiltext{1}{Division of Physics, Mathematics and Astronomy, 
                 California Institute of Technology, MS 59--33,
                 Pasadena, CA \ 91125}
\altaffiltext{2}{Current address: IPAC, 
                 California Institute of Technology, MS 100--22, 
                 Pasadena, CA \ 91125}
\altaffiltext{3}{Department of Physics,
                 Queen Mary and Westfield College,
                 Mile End Road,
                 London, E1 4NS, U.K.}
\altaffiltext{4}{Department of Physics,
                 University of California,
                 Berkeley, CA \ 94720}
\altaffiltext{5}{Current address:
                 Laboratory for Astrophysics and Space Research,
                 Enrico Fermi Institute,
                 University of Chicago,
                 5641 S. Ingleside Ave.,
                 Chicago, IL \ 60637}
\altaffiltext{6}{Current address: 
                 Department of Physics and Astronomy,
                 University of Massachusetts at Amherst,
                 Amherst, MA \ 01003--0120}
\altaffiltext{7}{Current address:
                 Aradigm Corporation,
                 26219 Eden Landing Road,
                 Hayward, CA \ 94545}

\begin{abstract}
  We present limits to anisotropies in the cosmic microwave background
  radiation~(CMB) at angular scales of a few arcminutes. The
  observations were made at a frequency of 142\,GHz using a 6-element
  bolometer array (the Sunyaev-Zel'dovich Infrared Experiment) at the
  Caltech Submillimeter Observatory.  Two patches of sky, each
  approximately $36'\times 4'$ and free of known sources, were
  observed for a total of 6-8 hours each, resulting in approximately
  80 independent $1\farcm7$ full-width half-maximum pixels.  Each
  pixel is observed with both a dual-beam and a
  triple-beam chop, with a sensitivity per pixel of 90-150\,$\mu$K in
  each chop.  These data have been analyzed using maximum likelihood
  techniques by assuming a gaussian autocorrelation function for the
  distribution of CMB fluctuations on the sky.  We set an upper limit
  of $\Delta T/T \le 2.1\times 10^{-5}$ (95\% confidence) for a
  coherence angle to the fluctuations of $1\farcm1$.  These limits are
  comparable to the best limits obtained from centimeter-wavelength
  observations on similar angular scales but have the advantage that
  the contribution from known point sources is negligible at these
  frequencies.  They are the most sensitive millimeter-wavelength
  limits for coherence angles $\le 3'$.  The results are also
  considered in the context of secondary sources of anisotropy,
  specifically the Sunyaev-Zel'dovich effect from galaxy clusters.
\end{abstract}


\section{Introduction}
\label{s1}
Measurements of anisotropies in the spatial distribution of the cosmic
microwave background radiation (CMB) are a powerful probe of the early
universe.  In the standard inflationary model, anisotropies were
imprinted on the CMB when the universe combined at $z\sim 1000$ (for a
review of cosmological theories see, for example, White, Scott and
Silk 1994). The attenuation of these primordial fluctuations in the
CMB by the finite thickness of the surface of last scattering, and
photon diffusion, suppress the anisotropy power spectrum at arcminute
angular scales.  The exact shape that is predicted for the power
spectrum in this regime is strongly dependent on the assumed
cosmological model.  Additionally, an early period of re-ionization
may have only a small effect on the anisotropy power spectrum at
degree scales while strongly affecting the magnitude and shape of the
power spectrum on arcminute scales.  Thus measurements of, or upper
limits on, the magnitude of the CMB power spectrum at arcminute scales
are a powerful discriminant between competing models.  Secondary
sources of arcminute-scale anisotropies include the Sunyaev-Zel'dovich
(S-Z) effect --- the inverse Compton scattering of CMB photons by hot
gas residing in the potential wells of galaxy clusters.  Measurements
at arcminute angular scales can provide important information for
theories of large-scale structure formation via number counts of
galaxy clusters with a measurable S-Z effect.

The Sunyaev-Zel'dovich Infrared Experiment (SuZIE) is a 6-element
bolometer array that has been used to make the first detections of the
S-Z effect at millimeter wavelengths (\cite{w3}, \cite{h1}). Coupled
to the Caltech Submillimeter Observatory (CSO), this array has a
sensitivity of 0.3\,Jy/$\sqrt{\mbox{Hz}}$ at 142\,GHz in each of six
pixel pairs and angular resolution of $\sim 1\farcm7$. This, and a scan
strategy designed to minimize systematic effects, make it an ideal
instrument with which to search for arcminute scale anisotropies in
the CMB.

This paper describes observations made with the SuZIE receiver of two
regions of sky that are free of known sources. The data are used to
set limits on CMB fluctuations that are assumed to be distributed on
the sky with a gaussian autocorrelation function (GACF).  A companion
paper (\cite{g1}) considers the application of these results to
specific cosmological models.  In this paper we also consider the
limits that these observations place on number counts of S-Z clusters.

\section{The Instrument}

The SuZIE bolometer array (described in detail in \cite{w4}) is
operated at the Caltech Submillimeter Observatory (CSO) on Mauna Kea.
At 142\,GHz, emission from the sky and the warm telescope optics are
measured by SuZIE to contribute 36\,K in brightness temperature.  A
tertiary mirror re-images the Cassegrain focus of the telescope to a
focal plane comprising two rows of three bolometers cooled to 300\,mK.
Each bolometer is coupled to the primary mirror by a Winston
concentrator that defines a $1\farcm7$ full-width half-maximum (FWHM)
beam on the sky.  The two rows are separated by $2'$, and adjacent
pixels within a row by $2\farcm3$.  The illumination pattern of each
Winston cone on the primary mirror is controlled by a 2\,K Lyot stop
located at an image of the primary produced by the tertiary mirror.
To reduce systematic effects from spillover, only 8\,m of the 10.4\,m
primary diameter are used.

SuZIE can be configured for operation at 142, 217 or 268\,GHz by
placing an appropriate metal-mesh band-pass filter in front of the
Winston cones.  The observations described here were made at 142\,GHz
with an 11\% band-width filter.  The spectral response of the
instrument, in its telescope-ready configuration, was measured in the
laboratory prior to the observations, including measurements that
demonstrate negligible out-of-band leaks.  The measured scatter of the
band centers of all six channels about the mean value is less than
1.5\% and that of the band-widths is less than 5\%.  The optical
efficiency of each channel, including the effects of all filters, is
measured to be 37\%.

Electronic differencing between pixels in the same row is carried out
to remove common-mode response to atmospheric and telescope emission.
Two differences corresponding to 2\farcm3 on the sky and one to
4\farcm6 are obtained from each row of the array.  This is
accomplished by placing the two bolometers in an AC-biased bridge
circuit (\cite{w2}), the output of which is synchronously demodulated
to produce a stable DC signal corresponding to the brightness
difference on the sky.  In terms of atmospheric subtraction, this is
equivalent to a square-wave chop on the sky at infinite frequency.
High rejection ($> 40\times$) of common-mode signals from detector
temperature fluctuations and telescope and atmospheric emission
(\cite{glezer1}) is achieved by trimming the amplitude of the bias
voltage, and thus the responsivity of each detector. To ensure that
the pixels that are being differenced are viewing the same column of
atmosphere, and thus that maximum rejection of common-mode atmospheric
fluctuations is obtained, the tertiary mirror is designed to maximize
the overlap of the beams from the six Winston cones on the primary
mirror without significantly degrading the focus.  The illumination of
the primary was measured at 217\,GHz using a mobile 80\,cm$^2$ load,
and was verified to fall to $<-60$\,db at a radius of 4.0\,m from the
primary center.

\section{The Observations}

SuZIE is designed to measure the S-Z effect in galaxy clusters.  As
part of this program, regions of sky free of known sources are
observed to provide a check for baseline effects that might
contaminate the S-Z measurements.  We selected two such regions for
observation in April 1994 at the positions listed in Table~\ref{t1},
using IRAS catalogs and the NRAO 5\,GHz survey (\cite{b2}) to avoid
known sources.  Each field is near a SuZIE target cluster, but at
least 10 core radii away from the cluster center.  The S-Z
contribution to each pixel from any residual hot gas at this distance
is $\Delta T_{\rm CMB} \le 3$\,$\mu$K, assuming an X-ray core radius
of $1'$, a $y$-parameter of $3 \times 10^{-4}$ and, pessimistically,
that the cluster gas is described by an isothermal, $\beta=2/3$, model
out to many core radii (see for example \cite{j1}).  Including the
effects of differencing pixels in the focal plane will reduce this
still further, depending on the angular scale on which the cluster gas
is clumped.

\placetable{t1}

\begin{figure}
\plotone{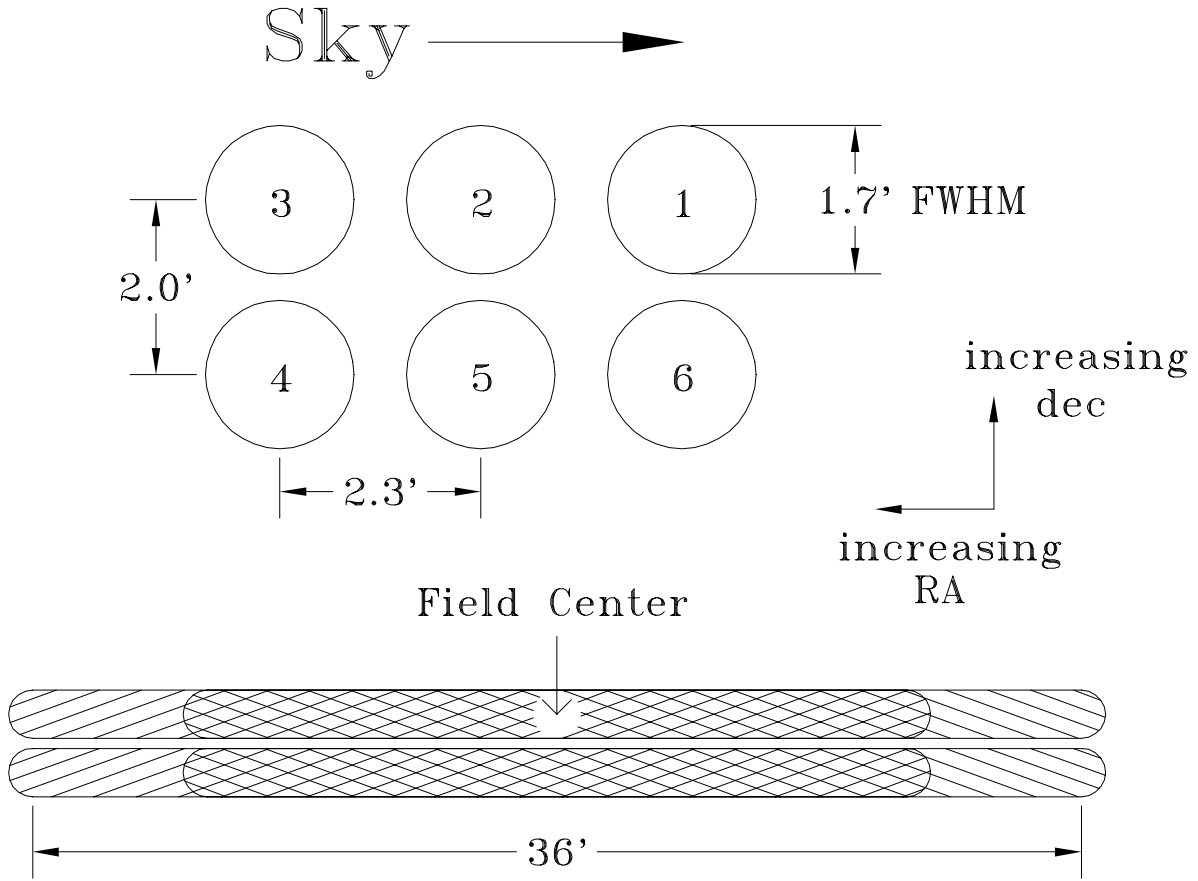}
\caption[]
{An illustration of the SuZIE scan strategy.  The upper portion shows
 the arrangement of the six pixels in the focal plane.  In the lower
 portion, the drift scan strategy is shown with the location of the
 field centers given in Table~1 indicated.  The opposing directions of
 the line hatching indicate the two RA offsets and the overlapping
 region is cross-hatched.}
\label{f1n} 
\end{figure}

In order to minimize systematic errors that might otherwise arise from
position-dependent variations in telescope spillover, SuZIE
observations are made by fixing the telescope position relative to the
earth and allowing the source to drift across the array as the sky
rotates.  Each scan is two minutes long, yielding a
$30'\times\cos\delta$ strip of sky, where $\delta$ is the source
declination. After each scan is completed, the dewar is rotated to
keep the long axis of the array aligned in the direction of right
ascension. The starting positions of successive scans alternate
between $12'$ and $18'$ in RA ahead of the location listed in
Table~\ref{t1}, providing a check for systematic effects that depend
on position within a scan rather than on the pointing location of the
telescope. Because there are two rows of detectors in the focal plane,
the complete data set for a single field covers two strips of sky each
$36'\cos\delta$ in RA by $1\farcm7$ in declination, with twice as much
observing time allotted to the central two-thirds of each strip as for
the outer portions.  The observing scheme is summarized in
Figure~\ref{f1n}.  Because both sources are close to the celestial
equator, $\cos\delta\sim 1$ and the total area of sky covered is 0.057
square degrees.

\placefigure{f1n}

\section{Data Reduction and Calibration}
\label{s4}

Seven data sets were obtained, each corresponding to one night of
observation on one field (typically 50-100 scans).  Each data set is
first split to separate the scans that correspond to the different RA
offsets.  The 6 differential measurements per data set that are
obtained by electronically differencing each detector with other
detectors in the same row are denoted by pairs $d_{pq}$ where $pq =$
12, 23 and 31 for the first row and 45, 56 and 64 for the second row;
the differential signal is the result of subtracting the $q$th from
the $p$th detector voltage.  The data were calibrated using drift-scan
observations of Uranus for which the expected flux was calculated by
convolving the spectral model for the source (\cite{g2}) with the
instrumental spectral response.  The uncertainty in
the absolute flux of Uranus at 142\,GHz obtained by this method is
$f_{\rm u}=\pm$6\%.  Corrections for atmospheric opacity due to the
difference between the elevation of Uranus and the elevation of the
Fields~1 and~2 observations were $< 2$\% (based on data from the CSO
225 GHz $\tau$-monitor extrapolated to 142\,GHz).  The result of the
calibration is then a signal whose units are flux difference between
the two beams.  This is converted to $\Delta T_{\rm CMB}$ (the
thermodynamic temperature difference between the two beams) as
follows:
\begin{equation}
  \Delta T_{\rm CMB} = \frac{\Delta I_{\rm CMB}}{I_{\rm CMB}}
  \frac{(e^{x}-1)}{xe^{x}} T_{\rm CMB}
\end{equation}
where $x = h\nu/kT_{\rm CMB}$ and:
\begin{equation}
  I_{\rm CMB} = \frac{2h \nu^3}{c^2}\frac{1}{(e^{x}-1)} \Omega_{\rm beam}
\end{equation}
The temperature of the CMB is taken to be 2.726\,K (\cite{m3}); the
effect of the 0.01\,K uncertainty in $T_{\rm CMB}$ is negligible
compared to other sources of calibration uncertainty.  

The solid angle of one pixel, $\Omega_{\rm beam}$, is calculated from
measurements of the beam profile made using Jupiter and Uranus.  The
2-dimensional shape of the beam was determined from drift scans across
Jupiter, with the array offset in declination from the source
(transverse to the scan direction) by steps of $15''$.  Jupiter is
used for this measurement because it is very bright at 142\,GHz and
can thus be used to obtain high signal/noise measurements of the
sidelobes and the aspect ratio of the beams.  However, its large size
($40''$) compared to the SuZIE beam makes it unsuitable for a direct
measurement of the beam solid angle.  This is determined instead from
drift scans across Uranus, by first assuming a circularly symmetric
beam on the sky and integrating the beam profile and then correcting
for the measured aspect ratio of the beams.  Note that the beams are
very nearly circular (see the Jupiter beam map in \cite{w4}) with an
aspect ratio (defined as FWHM in the scan direction divided by FWHM in
the cross-scan direction) that varies from 0.94--1.13.

The uncertainty in the beam shapes, $f_{\rm b}$, (estimated from the
scatter of the 6 measured solid angles about their mean value)
contributes a further 5\% to the calibration uncertainty.  The total
calibration uncertainty is then $f_{\rm c} = (f_u^2+f_b^2)^{1/2} =
8$\% ($1\sigma$).  This uncertainty is included in the likelihood
analysis of the data carried out in \S\ref{s7}.

Because each detector contributes to the signal measured by two pairs,
the six data streams are not independent.  To remove the degeneracy,
the two $2\farcm3$ pairs from each row are differenced to generate a
double difference:
\begin{eqnarray}
  t_{123} & = & (d_{12}-d_{23})/2 \\
  t_{456} & = & (d_{45}-d_{56})/2
\end{eqnarray}
The data then comprise 4 independent differences, two single
differences, $d_{31}$ and $d_{64}$, and two double differences,
$t_{123}$ and $t_{456}$.

Cosmic ray impacts on the bolometers cause spikes in the data stream
that must be removed.  A point-by-point differentiation is carried out
and spikes are identified by a large positive excursion followed by a
large negative excursion, or vice versa.  Approximately 5\% of the
data are removed by this process.  The data are then binned by
averaging together 15 samples (3\,s of data), equivalent to a
$0\farcm75\cos\delta$ portion of the drift scan, thus oversampling the
$1\farcm7$ beam full-width half-maximum (FWHM) by a factor of
$2.3/\!\cos\delta$.

A best fit offset and linear drift are removed from each scan prior to
co-adding all scans.  The linear drift arises from common-mode
variations in signal, primarily from changes in the atmospheric or
telescope temperature, that are not completely removed by
differencing.  The average values, and variance, of the removed linear
drift for each chop are shown in Table~\ref{t2}.  The effect of this
process on the sensitivity of the system to CMB fluctuations is small,
but is taken into account in the analysis in \S\ref{s7}.

\placetable{t2}

Because the residual noise is dominated by atmospheric emission
fluctuations with a $1\!/\!f$-type spectrum (\cite{w4}), the
statistical error obtained at each point from the binning process is
not a good estimate of the sensitivity of the system over longer
integration times.  The correlation time of the atmospheric noise can
be estimated from the correlation functions of individual scans and is
of order 5\,s (defined as the time to the first zero in the
correlation function).  Consequently, the binned data points within a
single scan are correlated over many bins.  The effects of this
correlation on the statistical properties of the data are considered
further in \S\ref{s6.3}.  In order to assign a sensible weight to each
scan, the rms scatter, $\sigma_j$, of the binned data in the $j$th
scan about the best fit offset and linear gradient is taken to be a
representative measure of the noise in that scan.  If the atmospheric
noise is normally-distributed within a scan and uncorrelated between
scans, this is a good assumption.  Note that this process also assumes
that the contribution to the rms from true temperature anisotropies
within a single scan is negligible.  The implications of these various
assumptions are discussed in
\S\ref{s6.3}.

\begin{figure}
\plotone{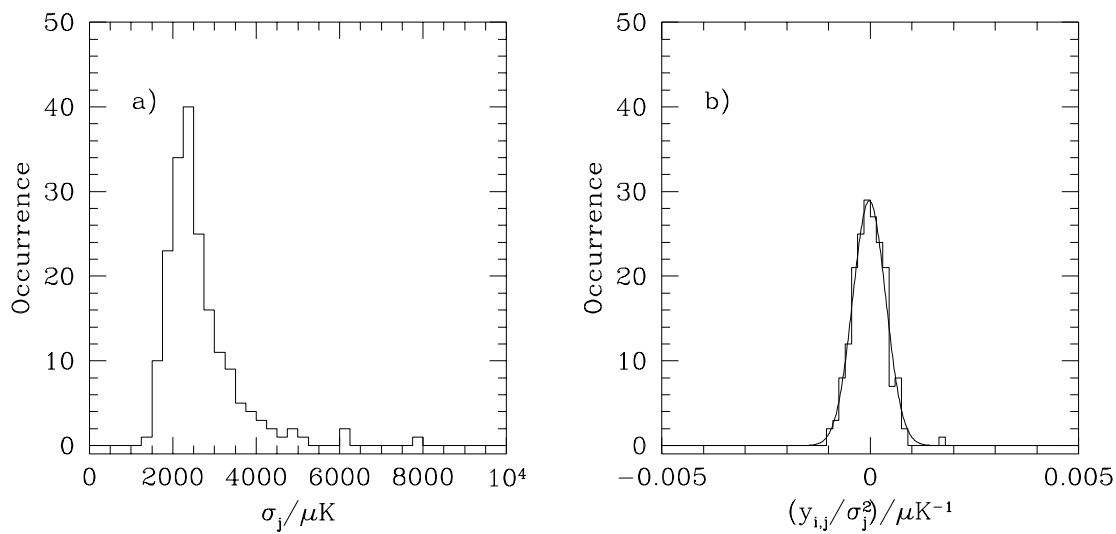}
\caption[]
{a) The distribution of the scan rms values, $\sigma_j$, for all 190
 scans obtained on Field~1 with the $d_{64}$ difference.  b) The
 distribution of the quantity $y_{i,j}/\sigma_j^2$ for a single bin
 from all 190 scans obtained on Field~1 with the $d_{64}$ difference.
 The solid line represents the normal distribution defined by the rms
 of $y_{i,j}/\sigma_j^2$ showing that this quantity is
 normally-distributed with no extreme values or large
 wings. }
\label{f2n} 
\end{figure}

The scans corresponding to a single RA offset observed on a single
night are co-added by combining all of the data at the $i$th point in
each scan to yield a signal $y_i$ with uncertainty $\sigma_i$
where:
\begin{eqnarray}
  y_i & = & \frac{\sum\limits_j y_{i,j}/\sigma_j^2}{\sum\limits_j 1/\sigma_j^2}
  \label{e1} \\ 
  \sigma_i^2 & = & \frac{1}{\sum\limits_j 1/\sigma_j^2} \hspace*{1cm}.
  \label{e2}
\end{eqnarray}
Here $\sigma_j$ is the rms of the $j$th scan after removal of the
offset and linear drift, as described above.  Note that
$\sigma_i=\sigma$, a constant across the entire co-added scan.
Figure~\ref{f2n}a shows the distribution of the $\sigma_j$ values for
Field~1 taken with the $d_{64}$ difference.  Note that {\em all} of
the data are included in the co-add and there is no cutting of data
based on high values of scan rms, $\sigma_j$.  Figure~\ref{f2n}b shows
the distribution of the quantity $y_{i,j}/\sigma_j^2$ evaluated for a
single bin, again using all of the Field~1 scans and data
corresponding to the $d_{64}$ difference.  It can be seen that this
quantity is normally-distributed with no extreme values or large
wings.  This indicates that the quantity $\sigma_j$ is a good estimate
of the variance across the entire data set.

\placefigure{f2n}

\begin{figure}
\plotone{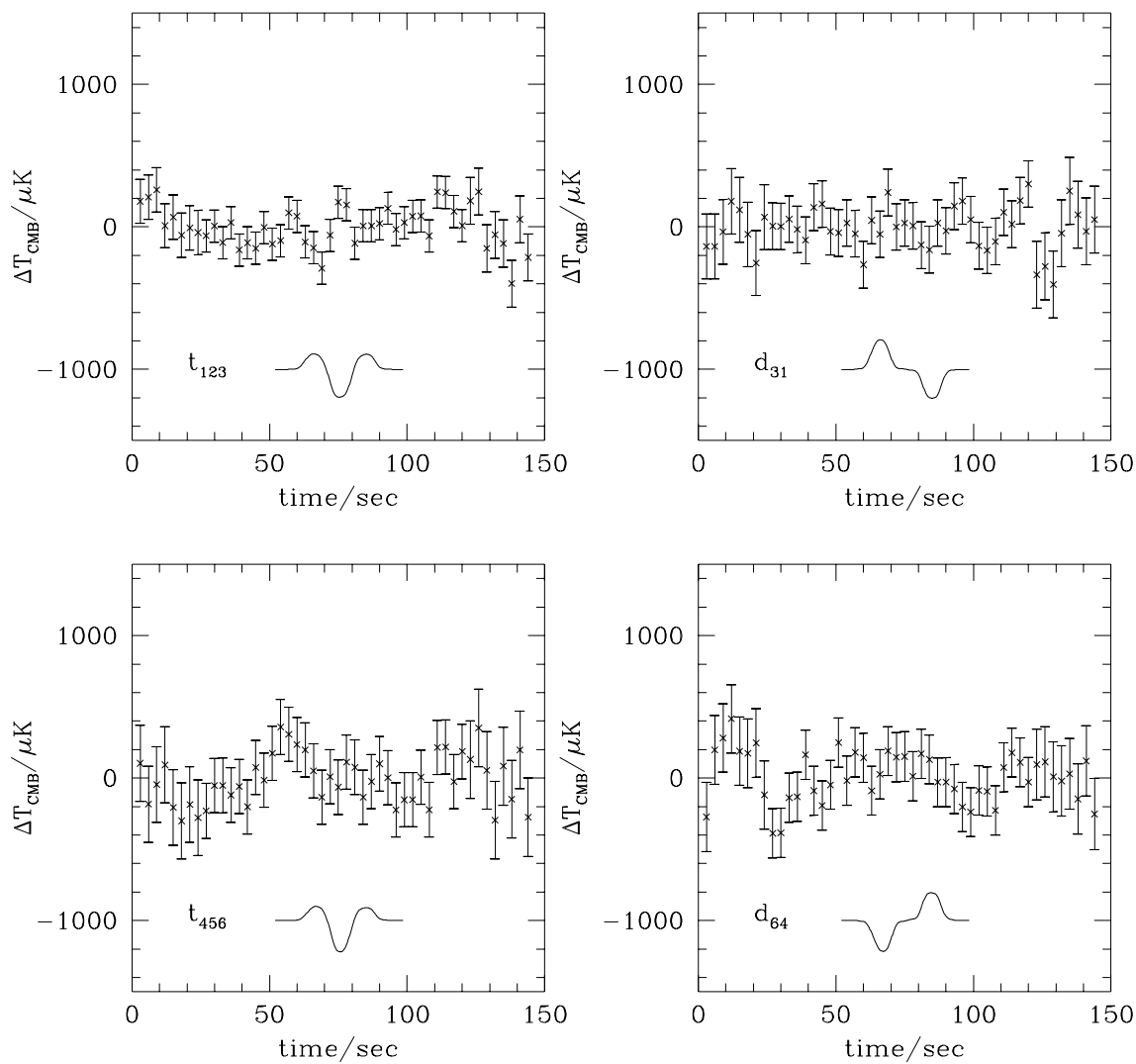}
\caption[]
{SuZIE measurements of Field~1.  The average uncertainty on each point
  is $\sim 130$\,$\mu$K per bin for $t_{123}$, $180$\,$\mu$K for
  $d_{31}$, 220\,$\mu$K for $t_{456}$ and 200\,$\mu$K for $d_{64}$.
  The instrument response to a point source is also shown in each
  panel. }
\label{f3n} 
\end{figure}

Co-added data from the two RA offsets are then combined by a weighted
average of data points that correspond to the same position on the
sky.  Finally, a weighted average of the co-added data sets from each
night is carried out.  Figure~\ref{f3n} shows the final co-added data
set for Field~1.  The rms uncertainty for each $0\farcm75$ bin is
130--220\,$\mu$K, corresponding to a sensitivity of
3--5\,mK\,s$^{1/2}$ for each difference.  Both of the single and
double difference data sets are shown, with the appropriate beam
response to a point source (measured by drift scans across Uranus)
indicated in each panel.  Approximately 6 hrs of integration time (190
scans) were obtained on Field~1 and 9 hrs (277 scans) on Field~2.
Note that double differencing reduces the error bars associated with
the $t_{123}$ data points relative to those of $d_{31}$ in
Figure~\ref{f3n}, as residual atmospheric gradients are being removed
from the data.  In the case of $t_{456}$ and $d_{64}$, no such
improvement is seen, probably due to poor common mode rejection in one
of the two pairs that are combined to form $t_{456}$.

\placefigure{f3n}

\section{Statistical Analysis of the Data}
\label{s5}
The data are first examined without prejudice as to the origin of any
possible excess signal by performing a series of statistical tests on
each data set.

The simplest test that can be performed on the data is a calculation
of the value of $\chi^2$ for each chop and each field, the results of
which are shown in Table~\ref{t2b}.  All chops have values of $\chi^2$
that lie within the 95\% ($\sim 2\sigma$) probability range for 46
degrees of freedom.  However there appears to be a systematic bias
towards $\chi^2$ values less than 46, indicating that the diagonal
correlation assumption is incorrect. This is the effect of the
residual correlation across the co-added data introduced by the
presence of atmospheric $1\!/\!f$ noise.  The implications of this
correlation are considered further is \S\ref{s6.3}.

\placetable{t2b}

A maximum likelihood analysis (\cite{la1}) has been carried out on
each data set to determine the most likely amplitude for any excess
variance above the noise term described by the uncertainties,
$\sigma_i$.  For simplicity, we assume that all points within a single
data set are independent, ignoring the correlation between adjacent
data points that is introduced by smearing of any true CMB signal by
the beams, or by the presence of $1\!/\!f$ noise in the data.

\begin{figure}
\plottwo{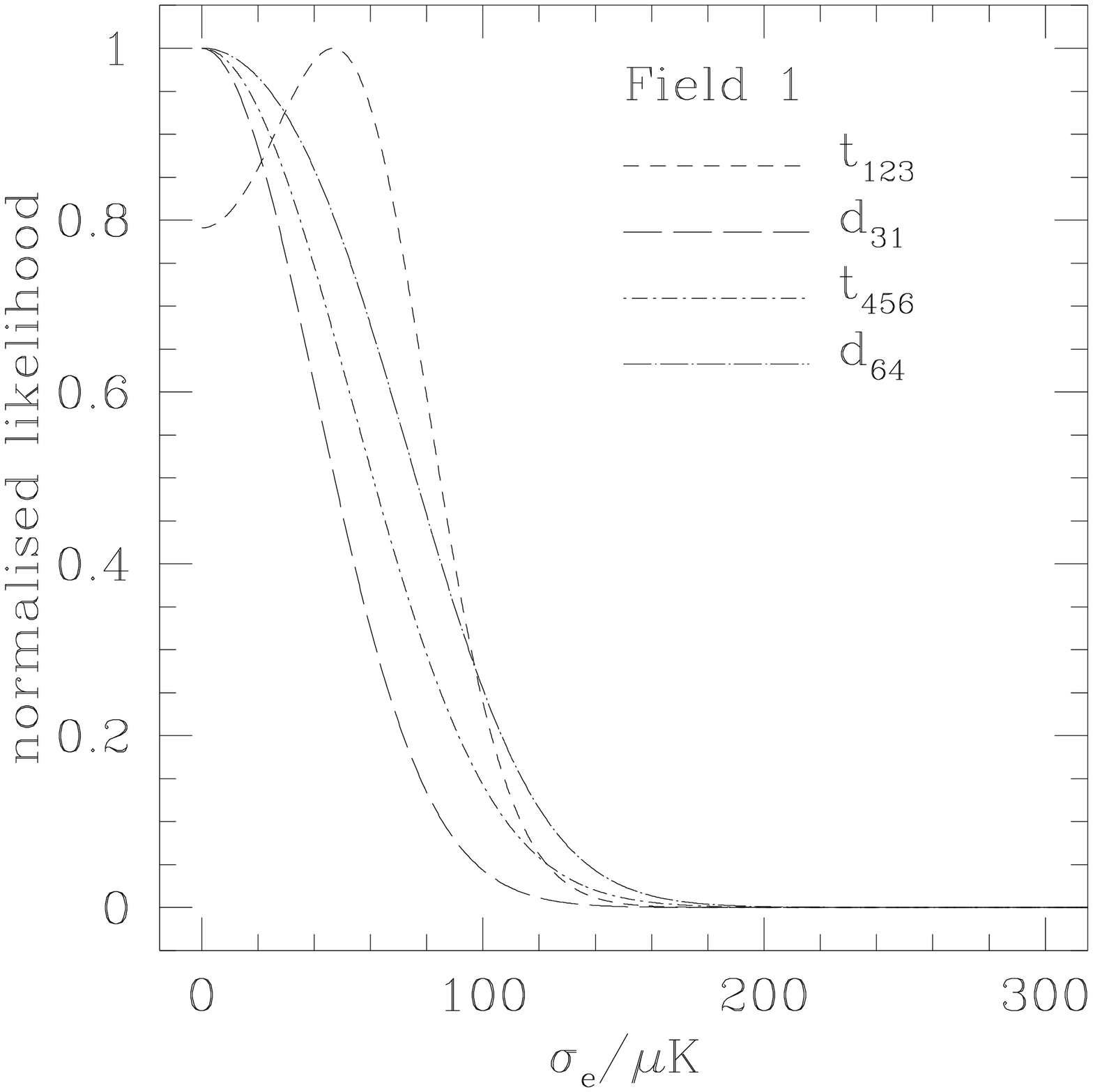}{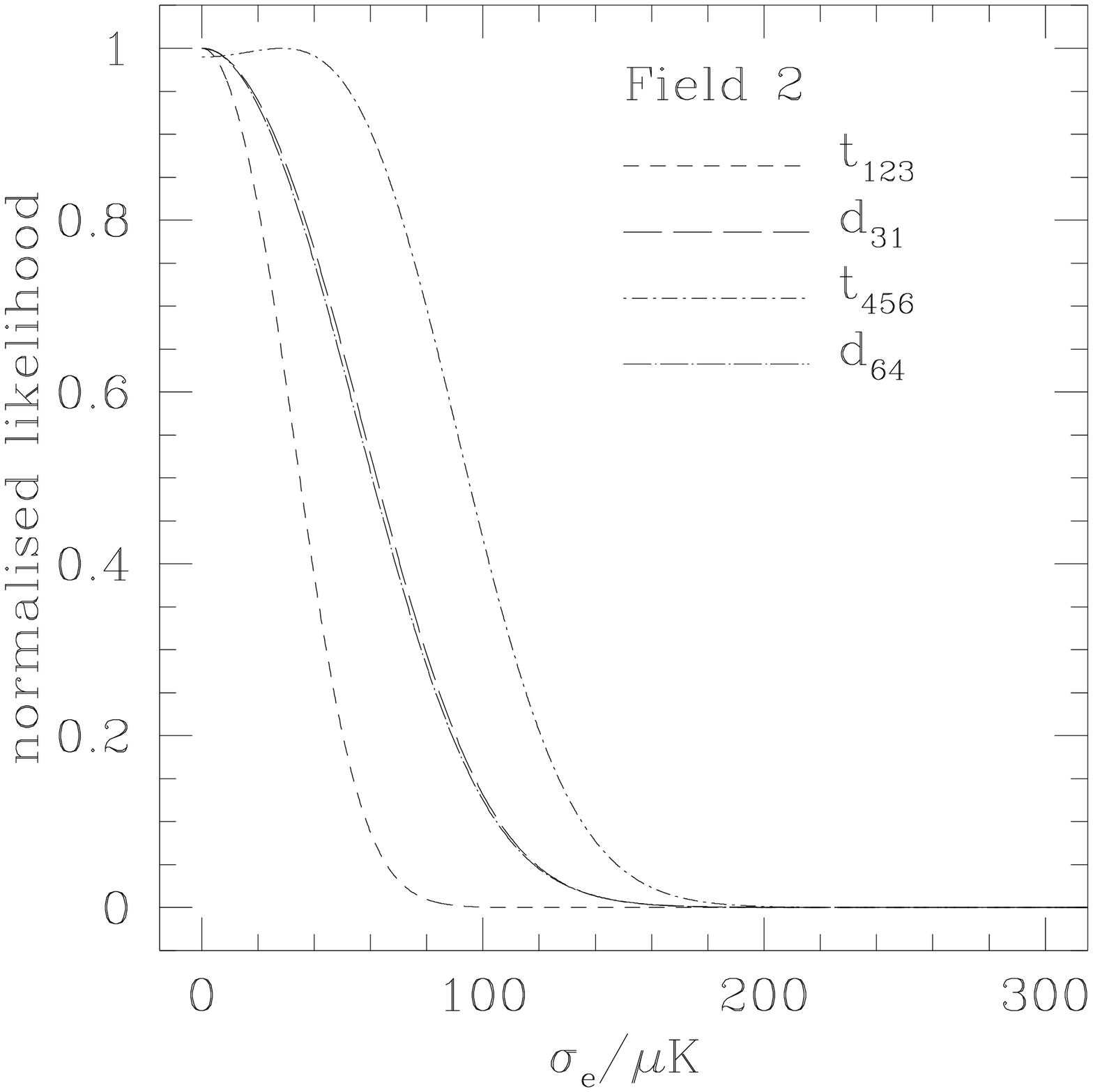}
\caption[]
{Likelihood values for excess variance, $\sigma_e$, in the SuZIE data,
  assuming no correlation between data points.  The likelihood
  function for each difference and each field is normalized to a peak
  value of 1. }
\label{f4n} 
\end{figure}

Each data set is analyzed separately using a maximum likelihood
estimator with the likelihood of an excess variance of $\sigma_e^2$
being:
\begin{equation}
  L(\sigma_e) = \prod\limits_{i=1}^{N}
  \frac{1}{[2\pi(\sigma_i^2+\sigma_e^2)]^{1/2}}
  \exp\left[\frac{-y^2_i}{2(\sigma_i^2+\sigma_e^2)}\right]
\end{equation}
The normalized likelihood as a function of $\sigma_e$ for each
difference within each field is shown in Figure~\ref{f4n}.  
There is no data set for which $\sigma_e\neq 0$ is
significantly more likely than $\sigma_e=0$, as expected from the
$\chi^2$ analysis.  To determine limits on $\sigma_e$ from these
curves, we adopt a Bayesian approach with a suitable choice of prior.
For consistency with experiments that span a similar range of angular
scales (e.g. \cite{m1}) we adopt a prior that is uniform in
$\sigma_e$.  The 95\% ($\sigma_{95} \sim 2\sigma$) and 99.7\%
($\sigma_{99.7} \sim 3\sigma$) limits are then calculated using the
highest probability density method (HPD, \cite{b10}), in which
\begin{equation}
  I = 100\times 
     \frac{\int_{\sigma_l}^{\sigma_u} L(\sigma_e) \, {\rm d}\sigma_e} 
                  {\int_0^\infty L(\sigma_e) d\sigma_e} ,
\end{equation}
with the constraint that $L(\sigma_l)=L(\sigma_u)$, yields $I=$ 95 or
99.7\%.  The values $\sigma_l$ and $\sigma_u$ are the lower and upper
confidence limits respectively.  The 95\% and 99.7\% confidence limits
to $\sigma_l$ are zero for all of the data sets presented here.  The
95\% and 99.7\% confidence limits for $\sigma_u$ are summarized in
Table~\ref{t3}.  The position of the likelihood peak is also
indicated.

%%% Ken's version
%\begin{equation}
%  I = 100\times { \int_0^{\sigma_I} L(\sigma_e) d\sigma_e \over
%                  \int_0^\infty     L(\sigma_e) d\sigma_e },
%\end{equation}
%where $I$ is either 95 or 99.7. These 95\% and 99.7\% confidence
%limits are summarized in Table~\ref{t3}.

\placefigure{f4n} \placetable{t3}

\begin{figure}
\plotone{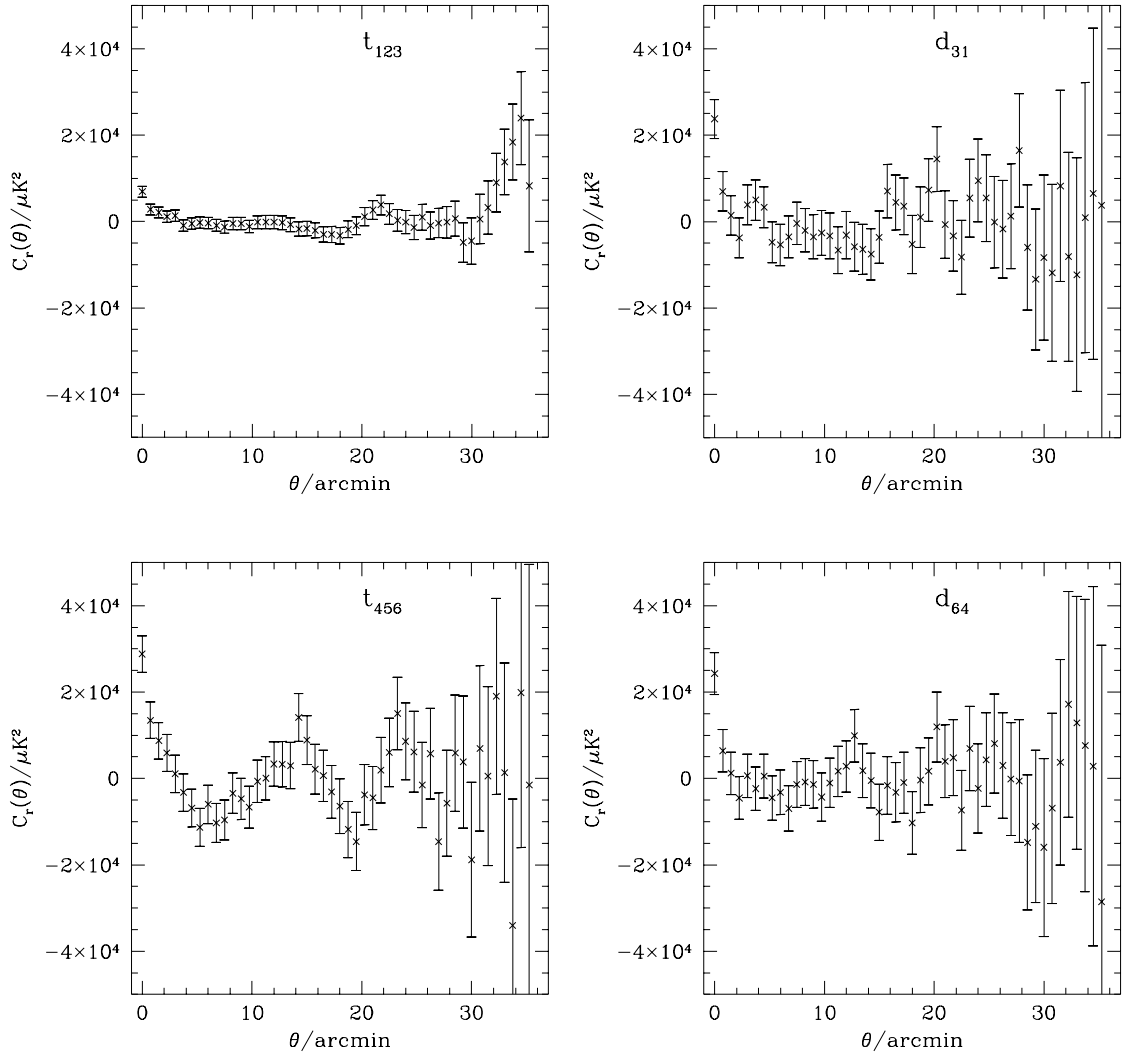}
\caption[]
{The correlation function of the co-added data from
 Field~2. }
\label{f5n} 
\end{figure}

We now examine the validity of the assumption that any excess variance
is uncorrelated across the data set.  The autocorrelation function of
the data, $C_r$, defined as:
\begin{eqnarray}
  C_r & = & \langle y_i \, y_{i+r} \rangle \\ 
      & = &  \frac{\sum\limits_{i=1}^{N-r} (y_i/\sigma_i^2)
        (y_{i+r}/\sigma_{i+r}^2)}{\sum\limits_{i=1}^{N-r}
        1/(\sigma_i^2 \, \sigma_{i+r}^2)}
  \label{e20}
\end{eqnarray}
(where $y_i$ and $\sigma_i$ are defined in equations~\ref{e1}
and~\ref{e2}) with an associated uncertainty, $\sigma_{C,r}$, given
by:
\begin{equation}
  \sigma^2_{C,r} = \frac{1}{\sum\limits_{i=1}^{N-r} 1/(\sigma_i^2 \,
    \sigma_{i+r}^2)}
\end{equation}
is shown in Figure~\ref{f5n} for data from Field~2.  Significant
structure can be seen in some of the differences.  Possible
non-astronomical sources for this structure are now considered.

\placefigure{f5n}

\section{Limits on Systematic Effects in the Data Sets}

\subsection{Systematic effects correlated with scan time}
\label{s6.1}
The standard SuZIE observing mode, in which the starting RAs of drift
scans are alternately 12$'$ and 18$'$ in RA ahead of the nominal
source position (see Figure~\ref{f1n}), allows checks for systematics
that are a function of time within a scan rather than pointing
location on the sky.  To check for such effects, the cross-correlation
of the co-added data corresponding to the 12$'$ offset with the
co-added data corresponding to the 18$'$ offset has been calculated
for each chop.  Artifacts that occur at the same time after the
beginning of a scan will be seen in the cross-correlation function as
a peak at zero time-lag.  True astronomical signals in the data will
yield a feature in the correlation function at a lag of $\Delta t =
-4\Delta\theta\cos\delta$ where $\Delta\theta=6'$ and $\delta$ is the
source declination.  Since both fields are close to the celestial
equator $\Delta t \approx -24$\,s.

\begin{figure}
\plotone{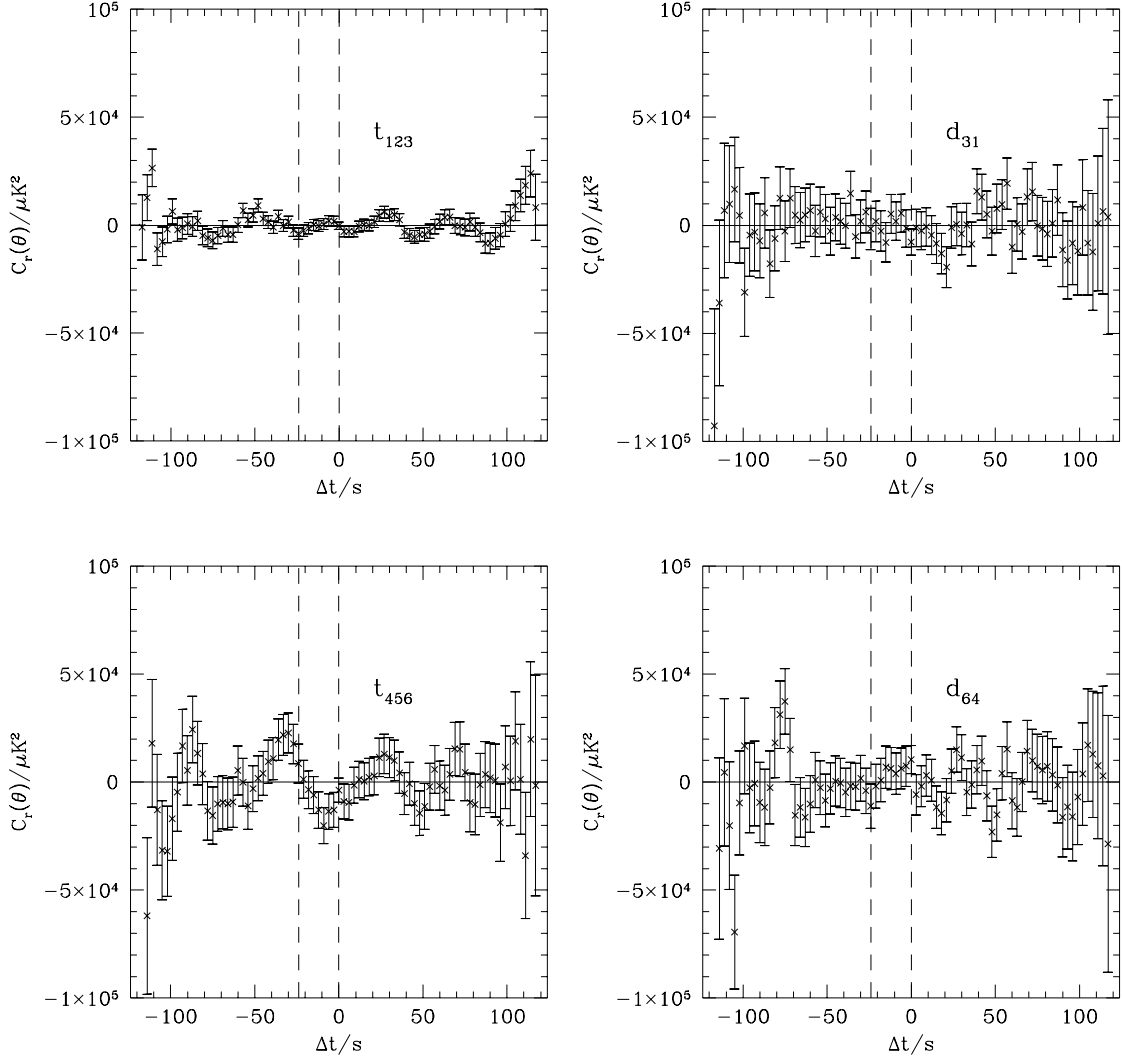}
\caption[]
{The cross-correlation of co-added scans corresponding to the 12$'$
  offset with co-added scans corresponding to the 18$'$ offset.  These
  data are taken from Field~2.  Systematic effects occurring at the
  same time after the beginning of a scan would yield a peak at
  $\Delta t=0$.  True astronomical signal would be seen as a peak at
  $\Delta t=-24$\,s.  These values are indicated by the dashed
  lines. }
\label{f6n} 
\end{figure}

Figure~\ref{f6n} shows the cross-correlation of data from the two
offsets for data sets corresponding to Field~2.  None of the chops
show any strong features at zero lag, indicating that there are no
systematic effects in the data that occur at the same time after the
beginning of a scan.  Peaks in the correlation functions at $\Delta t
\neq 0$ are seen, but there are no peaks at $\Delta t = -24$\,s that
would indicate the presence of true astronomical signal (the peak in
$t_{456}$ is at $-30$\,s and is not repeated in any other chop).  The
most likely cause of these peaks is chance correlation of
low-frequency $1\!/\!f$ noise which, in this experiment, is dominated
by atmospheric emission fluctuations (see \S\ref{s6.3}).  The long
($\sim 5$\,s) coherence time of atmospheric emission fluctuations
correlates data at the end of one scan with the data at the start of
the next.  The telescope moves by about $30'$ between scans,
corresponding to a change in position of the beams of $< 8$\,m at
heights $<1$\,km above the telescope.  This is less than the
correlation length of atmospheric structure which is $\sim 25$\,m
(assuming a wind speed of 5\,ms$^{-1}$ and a measured correlation time
of order 5\,s).  Co-adding many scans will reduce the magnitude of
this correlation but will not eliminate it entirely.  This effect
manifests itself in the cross-correlation functions shown in
Figure~\ref{f6n} as non-zero correlations observed at time lags of $\pm
120$\,s.

\placefigure{f6n}

\subsection{Cold stage temperature drifts}
\label{s6.2}

\begin{figure}
\plotone{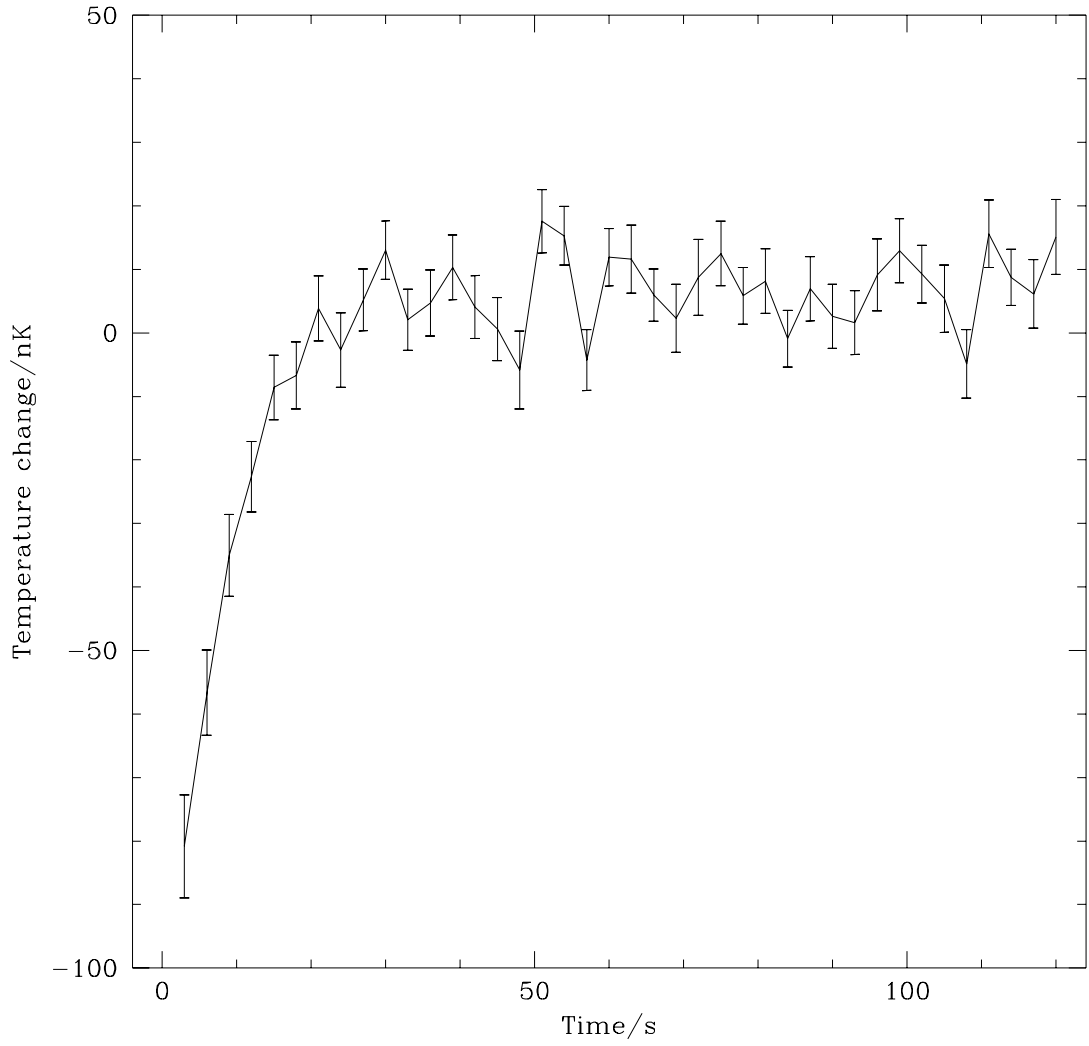}
\caption[]
{Co-added data from the 300\,mK stage temperature sensor showing the
  temperature excursion at the start of a scan induced by telescope
  motions. }
\label{f7n} 
\end{figure}

During SuZIE observations, the telescope is stationary during a scan,
then re-acquires the source before the next scan begins.  The motion
of the telescope between scans results in a small excursion in the
temperature of the 300\,mK stage at the start of a new scan.  The
output of the temperature sensor on the 300\,mK stage, co-added in the
same way as the science data, is shown in Figure~\ref{f7n}.  The
amplitude of the temperature excursion at the beginning of the scan is
$\sim 80$\,nK. The temperature recovers with a time constant of $\sim$
10~s, and is stable to $\sim$ 10~nK rms for the duration of the scan.

\placefigure{f7n}

Although the excursion at the beginning of each scan is very small, we
have carried out a number of checks to determine whether this
temperature drift causes residual artifacts in the co-added data sets.
First, as described in \S\ref{s6.1}, such an artifact would be seen as
a peak at zero lag in the cross-correlations of data from the two RA
offsets, shown in Figure~\ref{f6n}.  No significant peak is seen.
Second, since the temperature sensor data is sampled at an identical
rate to the science data, the measured temperature drift can be
correlated with the science channels on a scan-by-scan basis.  
% The average value of the temperature change in the science channels that
%is correlated to a change in temperature measured by the sensor is
%shown in Table~\ref{t2} and is $7-44$\,nK$_{\rm CMB}$ per nK change in
%temperature of the 300\,mK stage.
The maximum correlated change in signal in the science channels is
$\Delta T_{\rm CMB}<4$\,$\mu$K, well below other sources of noise.

\begin{figure}
\plotone{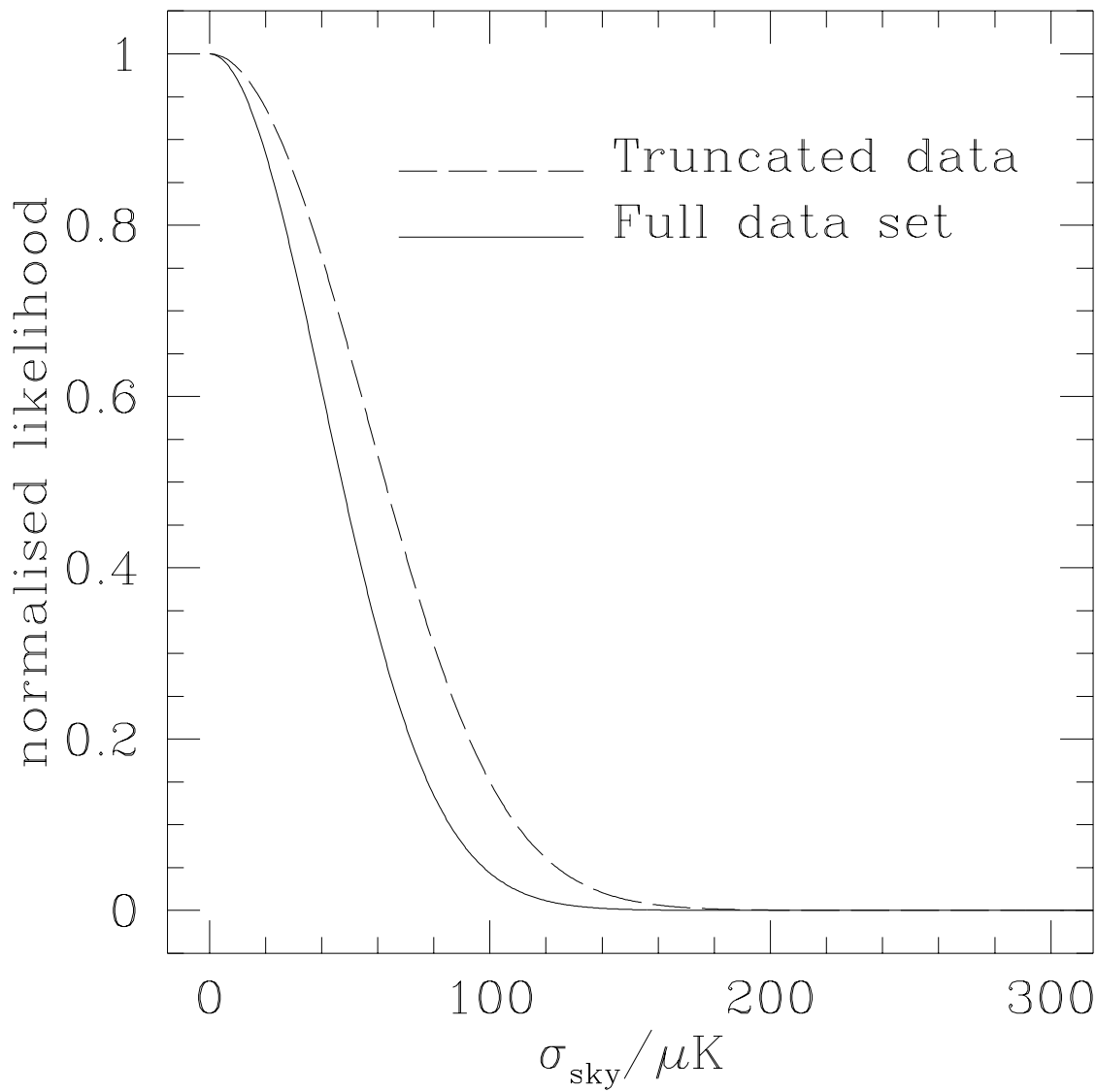}
\caption[]
{A comparison of the likelihood values for excess variance,
  $\sigma_e$, showing that when the first 21\,s of data from each scan
  are excluded from the analysis (dashed line) there is no significant
  change compared to the likelihood function for all the data
  (continuous line) other than that expected from loss of integration
  time.  These likelihood functions correspond to the $t_{123}$ chop
  from Field~1. }
\label{f8n}
\end{figure}

This analysis assumes that the shape and the phase of any temperature
change in the science channels is identical to that observed in the
temperature sensor.  Because of differences in the thermal properties
of the sensor and the bolometers, this may not be true.  Consequently,
there could be a large residual effect in the science data that is
poorly correlated to the sensor data.  As a final check, new co-added
data sets were generated after excluding the first 21\,s (7 bins) of
data from each scan, and the likelihood analysis in \S\ref{s5}
repeated.  As shown in Figure~\ref{f8n}, the sole effect on the
calculated likelihood is an increase in the confidence limits caused
by the effective reduction in integration time.

\placefigure{f8n}

From these various tests, we conclude that any signal in the data
caused by drifts in the temperature of the 300\,mK stage is much
smaller than the experimental uncertainties and can be ignored.

\subsection{Atmospheric effects}
\label{s6.3}

Fluctuations in atmospheric emission, caused predominantly by
variations in water vapor content, contribute $1\!/\!f$ noise to the
data with a typical correlation time of 5\,s, significantly less than
the scan length. We therefore assume that data corresponding to the
same coordinates on the sky but separated in time by more than one
scan are uncorrelated (this assumption ignores the correlation of data
at the end of one scan with data at the beginning of the next that
causes correlations on angular scales of $24'$ and $36'$ in the full
co-added data sets; we assume that the effects of atmospheric noise
correlated between scans are small). This assumption allows an
estimate of the residual contribution to the correlation function of
the co-added data.

Representing the $i$th data point in the $j$th scan by the sum of
$t_i$, the true astronomical signal that remains unchanged between
scans, and $n_{i,j}$, the noisy signal from all sources for which the
coherence time is less than the scan length, the correlation function
of the co-added signal is then:
\begin{equation}
  S_r = {\cal C}_r + {\cal N}_r
\end{equation}
Here ${\cal C}_r=\langle t_i \, t_{i+r}\rangle$ is the correlation function
of the true astronomical signal and ${\cal N}_r$ is the residual
correlation function of the noisy part of the signal.  It can be shown
(Appendix~A) that ${\cal N}_r$ is given by:
\begin{equation}
  {\cal N}_r = \frac{\sum\limits_j {\cal N}_{r,j}/\sigma^4_j}{(\sum\limits_j
    1/\sigma^2_j)^2}
  \label{e10}
\end{equation}
where ${\cal N}_{r,j}$ is the correlation function of $n_{i,j}$ in the
$j$th scan.  This expression can be rewritten as:
\begin{equation}
  {\cal N}_r =
  \left(\frac{\sum\limits_j {\cal N}_{r,j}/\sigma^4_j}{\sum\limits_j
      1/\sigma^4_j}\right) \times \left[\frac{\sum\limits_j
      1/\sigma^4_j}{(\sum\limits_j 1/\sigma^2_j)^2}\right]
\end{equation}
where the first term is the correlation function of the atmospheric
noise in a single scan averaged over all scans, and the second term is
equal to $1/N$ if the uncertainty, $\sigma_j$, is roughly constant
over all scans.  Since the correlation function of a single scan is
given by:
\begin{eqnarray}
  S_{r,j} & = & \langle(t_i + n_{i,j})(t_{i+r} + n_{i+r,j})\rangle\\
          & = &  {\cal C}_r + {\cal N}_{r,j} 
\end{eqnarray}
it is clearly not possible to calculate ${\cal N}_{r,j}$ independently
of the contribution from the true astronomical correlation function.
However, if we assume that the atmospheric noise within the individual
scans dominates the true astronomical signal in each scan then ${\cal
  N}_{r,j} \sim S_{r,j}$ and the estimated residual correlation
function from atmospheric noise, ${\cal N}^e_r$, can then be
calculated using Equation~\ref{e10} with ${\cal N}_{r,j} = S_{r,j}$.
Note that ${\cal N}_{r,j} \gg {\cal C}_r$ is implicit in the
assumption made in \S\ref{s4}, Equations~\ref{e1} and~\ref{e2}, that the
uncertainty, $\sigma_{i,j}$, associated with the data points in an
individual scan is equal to $\sigma_j$, the rms of an entire scan.

How valid is this assumption?  Since the residual atmospheric noise
calculated using Equation~\ref{e10} is approximately equal to the
average correlation function per scan divided by the total number of
scans, the contribution of the true astronomical correlation function,
${\cal C}_r$, to ${\cal N}^e_r$ calculated using Equation~\ref{e10} is
$\sim {\cal C}_r/N$.  Since N is typically 100 per RA offset for each
field, this contribution is clearly very small compared to ${\cal
  C}_r$.  Therefore, if the calculated function, ${\cal N}^e_r$, is
compared with the measured correlation function of the co-added data
there are several possibilities: (i) if ${\cal N}_{r,j}$ and ${\cal
  N}_r \gg {\cal C}_r$ then the contribution of ${\cal C}_r$ to ${\cal
  N}^e_r$ will be negligible and ${\cal N}^e_r$ will be a good fit to
the correlation functions shown in Figure~\ref{f5n}; (ii) if ${\cal
  N}_{r,j}$ and ${\cal N}_r \ll {\cal C}_r$ then the calculated
function ${\cal N}^e_r$ will be $\sim {\cal C}_r/N$.  Thus ${\cal
  N}^e_r$ will have the right shape, but will have an amplitude that
is too low by approximately two orders of magnitude; (iii) if ${\cal
  N}_r$ and ${\cal C}_r$ have comparable magnitudes then ${\cal
  N}^e_r$ will be a poor fit to the measured correlation functions.

\begin{figure}
\plotone{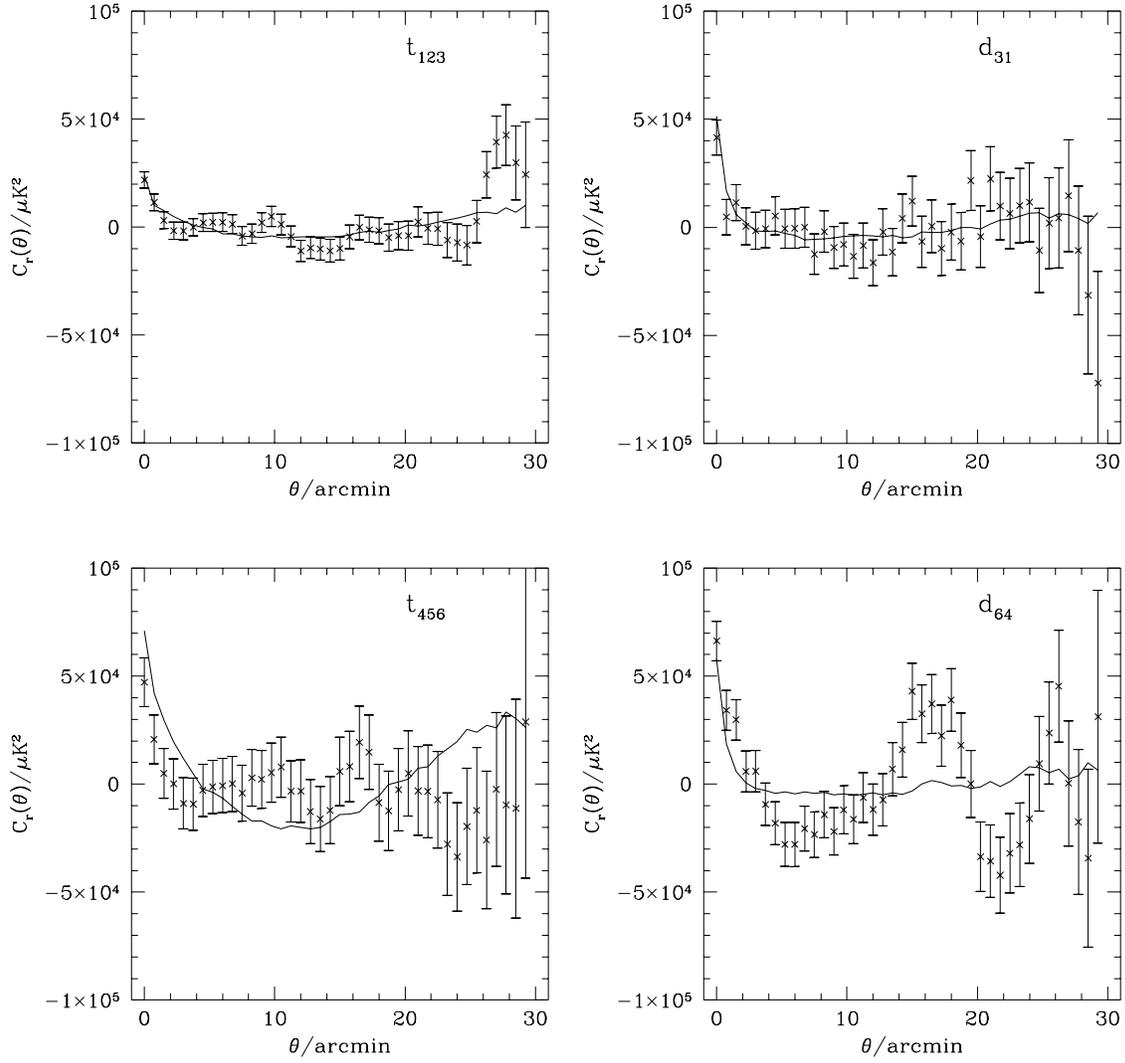}
\caption[]
{Correlation function of one RA offset of co-added data from Field~1.
  The continuous line is the residual correlation function calculated
  using Equation~\protect\ref{e10}. }
\label{f9n} 
\end{figure}

Figure~\ref{f9n} shows the estimated correlation function of the
residual atmospheric noise, ${\cal N}^e_r$, calculated using the
methods given above, overlaid on the correlation function of the
co-added data corresponding to one RA offset taken from Field~1.  At
angles less than $10'$, the correlation function of the co-added data
is moderately well fitted by ${\cal N}^e_r$, corresponding to case~(i)
above and indicating that the majority of the observed correlation in
the data on these scales is due to atmospheric noise or other sources
of $1\!/\!f$ noise.  There are some large deviations however,
particularly in the data from row~2 (pixels 4, 5 and 6).  Since this
row is known to be more susceptible than row~1 to atmospheric
fluctuations, these large correlations are likely to be the result of
atmospheric drifts that are correlated on time scales longer than one
scan (see \S\ref{s6.1}).

\placefigure{f9n}

\section{Comparison of the Data with Gaussian Models of CMB Anisotropies}
\label{s7}
In order to compare the data to specific models for the power spectrum
of CMB fluctuations, the correlation properties introduced by the beam
response on the sky, and the overlap on the sky of the single and
double differences, must be taken into account.  We represent the
correlation function of true temperature anisotropies by:
\begin{equation}
  {\cal C}_r = C({\bf n}_a {\bf n}_b) = \langle T({\bf n}_a)T({\bf
    n}_b) \rangle
\end{equation}
where ${\bf n}_a$ and ${\bf n}_b$ are directions on the sky.  If the
CMB sky is assumed to be sampled from a gaussian random field then
$C({\bf n}_a \, {\bf n}_b)$ can be written as $\Delta T_0^2 \times
{\cal F}(\theta)$ where the spatial dependence of the function ${\cal
  F}(\theta)$ depends only the angular separation, $\theta$, between
the two vectors ${\bf n}_a$ and ${\bf n}_b$.  The model correlation
function, $\overline{C}_r$, of the measured data in the absence of any
noise sources is then given by:
\begin{equation}
  {\cal C}_r = \overline{C}_r = \langle \overline{T}_i \,
  \overline{T}_{i+r} \rangle = \Delta T_0^2 \times \overline{\cal F}
  (r\Delta\theta)
\label{e7.1}
\end{equation}
where $\overline{T}_i$ and $\overline{T}_{i+r}$ are the predicted
thermodynamic temperatures of pixels $i$ and $i+r$ in the data stream
after convolution of the true temperatures with the beam of the
instrument.  The quantity $\Delta\theta$ is the sampling rate on the
sky. For instance, the correlation function between data points
measured in the single difference data set, $d_{31}$, is given by:
\begin{equation}
  \langle \overline{T}_i \, \overline{T}_{i+r} \rangle_{31} = 
  \int\limits_{{\bf n}_a} \int\limits_{{\bf n}_b} 
  \langle T({\bf n}_a) \, T({\bf n}_b) \rangle \
  [A_3({\bf n}_i-{\bf n}_a)-A_1({\bf n}_i-{\bf n}_a)]  \
  [A_3({\bf n}_{i+r}-{\bf n}_b)-A_1({\bf n}_{i+r}-{\bf n}_b)] \
  {\rm d}{\bf n}_a {\rm d}{\bf n}_b
\label{e7.5}
\end{equation}
where $A_j$ is the beam response function of pixel $j$ in the SuZIE
array.

A simultaneous likelihood analysis for all data from one field can be
performed by combining all four differences into one data vector ${\bf
  y}$ of length $4N$, where $N$ is the number of 45$''$ data bins. The
likelihood of the model correlation function, given the data, can be
calculated as follows:
\begin{equation}
  L(\Delta T_0,{\cal F}) =
  \frac{1}{(2\pi)^{N/2}|{\bf M}|^{1/2}} \exp(-{\bf y}^T{\bf M}^{-1}\,{\bf y})
  \label{e7.2}
\end{equation}
The $4N \times 4N$ matrix ${\bf M}$ is composed of sub-matrices ${\bf
  m}^{pq}$ where $p,q=1,4$.  The elements of each sub-matrix are given
by:
\begin{equation}
  m^{pq}_{i,j} = \overline{C}^{pq}_{j-i} + {\cal N}^{pq}_{j-i}
  \label{e7.4}
\end{equation}
where $i,j=1,N$.  The function $\overline{C}^{pq}_{j-i}$ is given by
Equation~\ref{e7.5} with the beam responses for the appropriate
differences inserted.  The function ${\cal N}^{pq}_{j-i}$ is the
correlation function of the noise component to the data.

We need, however, to account for the loss of degrees of freedom
introduced by the removal of a mean and linear drift from each scan.
The process, adapted from Bond et al.\ (1991), is summarized here.
The true CMB signal that has been subtracted from the data by the
removal of an offset and linear drift can be represented as an unknown
amount, ${\bf a}$, of a set of fitting functions, ${\bf F}$.  The data
vector, ${\bf y}$, can be corrected for this process by defining a
vector ${\bf y}_{\rm true}$ such that:
\begin{equation}
  {\bf y} = {\bf y}_{\rm true} - {\bf a}^T{\bf F}
\end{equation}
The unknown vector ${\bf a}^T$ has length 8 (the product of the number
of differences and two fitting functions); ${\bf F}$ is a known
$8 \times 4N$ matrix with elements equal to the value of the fitting
functions ($F_{i,j}=1$ for $i=1,3,5,7$ is the basis function for the
offset, and $F_{i,j} = j$ modulo $N$ for $i=2,4,6,8$ is the basis
function for the slope, where $j=1,4N$).  Thus the likelihood function
in Equation~\ref{e7.2} should be written as:
\begin{equation}
  L(\Delta T_0,{\cal F}) = 
  \frac{1}{(2\pi)^{N/2}|{\bf M}|^{1/2}} \exp[-({\bf y}+{\bf a}{\bf
    F})^T{\bf M}^{-1}\,({\bf y}+{\bf a}{\bf F})]
\end{equation}
Assuming a uniform prior for, and integrating over, the unknown vector ${\bf a}$ yields:
\begin{equation}
  L_{\rm new} (\Delta T_0,{\cal F}) \propto \frac{1}{(|{\bf M}||{\bf F}^T{\bf
      M}^{-1}{\bf F}|)^{1/2}} \exp\left[\frac{-{\bf y}^T {\bf M}^{-1} {\bf
      y} + {\bf y}^T{\bf M}^{-1}{\bf F}({\bf F}^T{\bf M}^{-1}{\bf
      F})^{-1}{\bf F}^T ({\bf M}^{-1})^T{\bf y}}{2}\right]
  \label{e7.3}
\end{equation}

Extending this analysis to include data from both fields is quite
simple since the two patches of sky are widely separated on the sky,
and also widely separated in terms of observing time.  Thus the two
data sets are completely uncorrelated both spatially and temporally
and the matrix ${\bf M}$ for the complete data set is:
\begin{equation}
{\bf M}_{\rm tot} = \left(  \begin{array}{cc}
                                 {\bf M}_1 & 0 \\
                                       0   & {\bf M}_2 \\
                            \end{array}
                    \right)
\end{equation}
The likelihood function for all data from both fields is then:
\begin{equation}
  L_{\rm tot} =L_1 \times L_2
  \label{e21}
\end{equation}
where $L_1$ and $L_2$ are calculated from Equation~\ref{e7.3}.  The
calibration uncertainty is included at this point in the following
manner.  The measured amplitude of the sky correlation function,
$\Delta T'_0$ is assumed to be related to the true value $\Delta T_0$
by $\Delta T'_0= G \times \Delta T_0$, where G is gaussian-distributed
with width $\sigma_G$ (the fractional calibration uncertainty).  It
can then be shown (\cite{g4}) that the likelihood of the true
amplitude is obtained from the likelihood of the measured amplitude
by:
\begin{equation}
  L(\Delta T_0,{\cal F}) = \frac{1}{\sqrt{2\pi} \,\sigma_G \, \Delta T_0} \times
  \int\limits_{0}^{\infty} d(\Delta T'_0)
  \exp\left[ \frac{-(\Delta T'_0-\Delta T_0)^2}{2\sigma_G^2\Delta T_0}
\right]
L(\Delta T'_0,{\cal F}) 
\end{equation} 

There remains the issue of the noise term ${\cal N}^{pq}_{j-i}$ in
Equation~\ref{e7.4}.  The simplest assumption is that the noise terms
are uncorrelated between separate pixels and differences and that
${\cal N}^{pq}_{j-i} = \delta(p-q,j-i) \times (\sigma^p_i)^2$ where
$\sigma_i^p$ is the uncertainty associated with the $i$th data point
in difference $p$.  However, as shown in \S\ref{s6.3}, there is a
significant component of $1\!/\!f$ noise in the data that introduces
correlations between pixels and also between the different chops.  In
principle, the non-diagonal noise terms can be estimated using
Equation~\ref{e10}.  However, this will not account for the residual
features in Figure~\ref{f9n} that are believed to be caused by slow
drifts correlated across more than one scan, and that are not well
fitted by Equation~\ref{e10}.

Instead we adopt the following approach.  Because ${\cal N}^e_r$ is
such a good fit to the correlation function of the co-added data,
there is unlikely to be a high signal/noise detection of astronomical
signal.  Thus the data can be analyzed assuming the simple
$\delta$-function form for ${\cal N}^{pq}_{j-i}$ with the constraint that the
results from the likelihood analysis should be treated as upper limits
only to any signal in the data.  The possible effects of ignoring the
correlated noise component are discussed in \S\ref{s7.1}.

To summarize, the elements of the sub-matrices ${\bf m}^{pq}$ are
$m^{pq}_{i,j} = \overline{C}^{pq}(r\Delta\theta) +
\mbox{$\delta(p-q,r)$} \, (\sigma_p)^2$ where $r=j-i$ and
$\Delta\theta=0\farcm75\times\cos\delta$ is the sampling interval of
the data on the sky.  The modified correlation function,
${\overline{C}}^{pq}(r\Delta\theta)$ is obtained from
Equation~\ref{e7.5}.

\subsection{Comparison With a Gaussian Autocorrelation Function}
\label{s7.1}
The generalized form for $C(\theta)$ is an expansion in terms of
Legendre polynomials, $P_l(\cos\theta)$:
\begin{equation}
C(\theta) = \frac{1}{4\pi} \sum\limits_{l=2}^{\infty} (2l+1) C_l
P_l(\cos\theta)
\end{equation}
where the function $C_l$ depends on the assumed cosmological model.
The functional form of $C_l$ due to anisotropies imprinted at
recombination has been extensively modeled in the literature
(\cite{w1}, \cite{b8}, \cite{g3}, \cite{r4}) and the
implications of the SuZIE measurements for these models are considered
in a companion paper (\cite{g1}).  In this paper we consider our
measurements in terms of secondary sources of anisotropy from a
population of S-Z clusters, since these may be comparable to, or
larger than, primary anisotropies at these angular scales (\cite{b4}).
The analytically simple gaussian autocorrelation function (GACF) model
for $C(\theta)$, while a poor approximation to the power spectrum of
primordial anisotropies, is a useful tool in this case.

The functional forms for the GACF is:
\begin{equation}
  C(\theta) = C_0 \exp\left[\frac{-\theta^2}{2\theta_0^2}\right]\\
\end{equation}
with two free parameters, $C_0=\Delta T^2_0$, the variance, and
$\theta_0$, the coherence angle, of the fluctuations.  In the absence
of other noise sources, observations of anisotropies distributed with
this correlation function, using a gaussian beam of FWHM $\theta_b$,
yield measurements with correlation function:
\begin{equation}
  \overline{C}(\theta) = \frac{C_0\theta_0^2}{2\beta^2+\theta_0^2} 
  \exp\left[\frac{-\theta^2}{2(2\beta^2+\theta_0^2)}\right]
\end{equation}
where $\beta=\theta_b/\sqrt{8\log2}$.

The SuZIE beams are assumed to be gaussian, with a FWHM of 1\farcm7,
derived by normalization to the measured solid angle.  If a gaussian
is fitted directly to the beam profiles, then the derived FWHM is 6\%
larger, however, this gaussian overestimates the beam shape at several
beam widths away from the beam center.  The 6\% difference can be
taken as some measure of the uncertainty that is introduced by
assuming a gaussian beam.  Ganga et al.\ (1997b) show that beam
uncertainties of this magnitude do not affect the results of the
modeling although they must be included in the calibration
uncertainty, as we have done in \S\ref{s4}.

\begin{figure}
\plotone{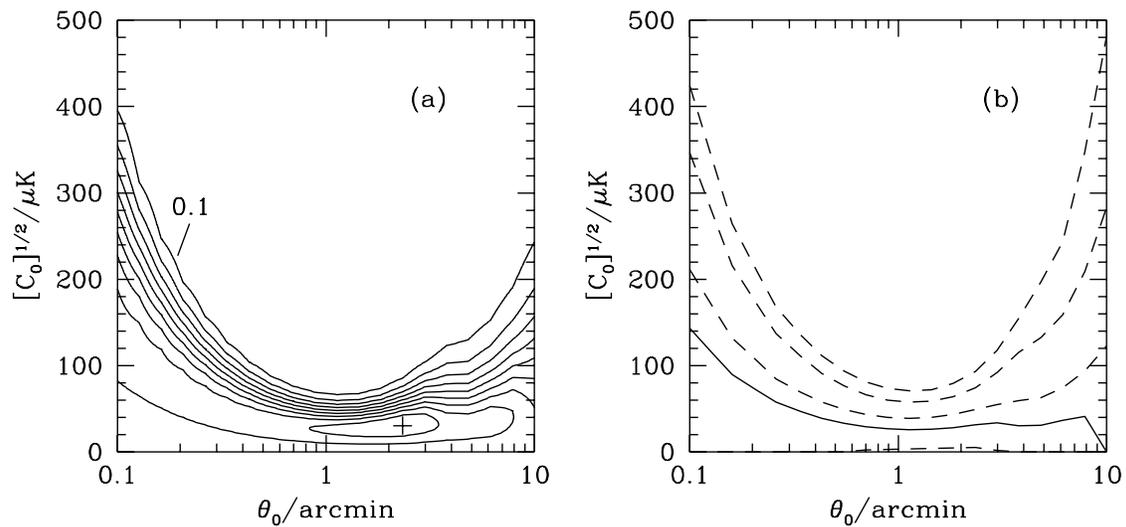}
\caption[]
{a) Likelihood contours, normalized to a peak likelihood of 1,
  obtained by assuming a gaussian autocorrelation function for the
  distribution of CMB anisotropies.  Contour levels start at 0.1
  (labeled) and increase in steps of 0.1.  The most likely value of
  $C_0^{1/2}$ and $\theta_0$ is marked with a cross.  b) The
  continuous line shows the position of the maximum likelihood for
  each value of $\theta_0$, the dashed lines show 63\%, 95\% and
  99.7\% confidence limits. }
\label{f10n} 
\end{figure}

The likelihood contours as a function of $(C_0)^{1/2}$ and coherence
angle, $\theta_0$, from a simultaneous likelihood analysis of all data
from all fields, are shown in Figure~\ref{f10n}a.  A peak in the
likelihood function at non-zero values of $C_0\neq 0$ is obtained for
all coherence angles less than $10'$.  Using the HPD method with a
uniform prior in $(C_0)^{1/2}$, the 68\%, 95\% and 99.7\% confidence
limits on $(C_0)^{1/2}$ for each value of $\theta_0$ have been
calculated, and are shown in Figure~\ref{f10n}b.  The peak in the
likelihood function has low significant ($\le$ 68\% confidence) and is
consistent with noise.  Consequently we confine ourselves to
considering the 95\% and 99.7\% confidence limits.  The coherence
angle at which maximum sensitivity to $C_0$ is obtained is $\theta_0 =
1\farcm1$; at this coherence angle, $(C_0)^{1/2} \le$~58\,$\mu$K (95\%
confidence) and $\le$~72\,$\mu$K (99.7\% confidence).  Expressing
these limits in terms of $\Delta T/T$ yields:
\begin{equation}
\Delta T/T \le \left\{  \begin{array}{ll}
   2.1\times 10^{-5} \mbox{\hspace*{5mm} (95\% confidence)}\\
   2.6\times 10^{-5} \mbox{\hspace*{5mm} (99.7\% confidence)}
                        \end{array}
               \right.
\label{e31}
\end{equation}

\placefigure{f10n}

These limits have been calculated without including the correlated
terms in the matrix {\bf M} that are introduced by the presence of
atmospheric noise in the data.  As such we believe the limits to be
conservative.  To check this assumption, we have re-analyzed just the
data corresponding to the $d_{31}$ and $d_{64}$ chops, at the
coherence angle of maximum sensitivity ($\theta_0=1\farcm1$) only,
with the inclusion of the terms ${\cal N}^{pq}_{j-i}$ in
Equation~\ref{e7.4}.  These terms have been calculated as outlined in
\S\ref{s6.3}.  Ignoring data from the triple beam differences,
$t_{123}$ and $t_{456}$, does not significantly degrade the derived
upper limits.  We find that the limits that are derived when the
correlated noise terms are included are somewhat lower than those
calculated when the correlated terms are ignored.  Consequently the
assumption that our upper limits are conservative is valid.

\begin{figure}
\plotone{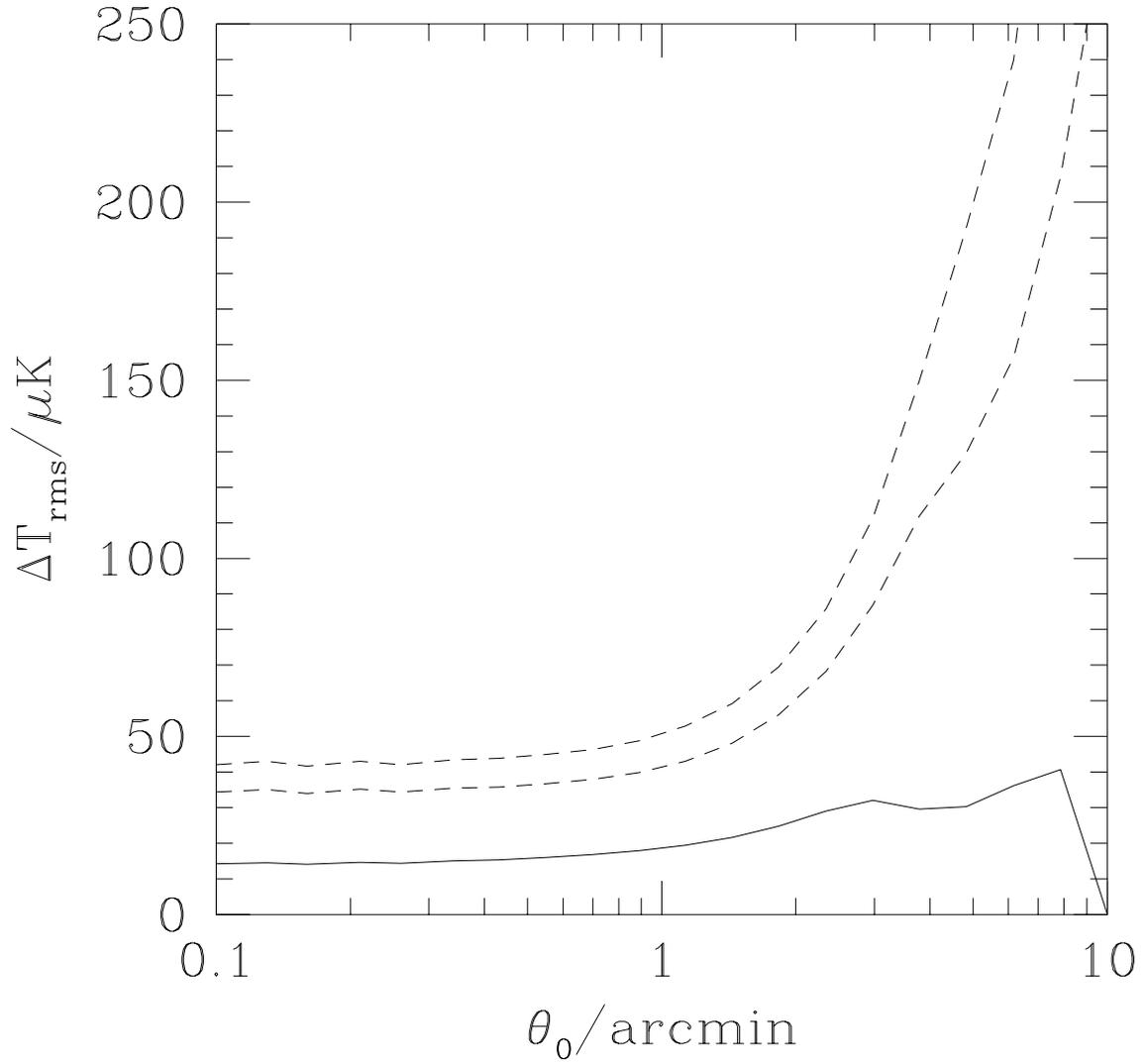}
\caption[]
{Limits on $\Delta T_{\rm rms} = [\overline{C}(0)]^{1/2}$ as a
  function of coherence angle, $\theta_0$.  The bold line shows the
  most likely value, the dashed lines show the 95\% and 99.7\% upper
  limits. }
\label{f11n}  
\end{figure}

It is also useful to express these limits in terms of $\overline{C}(0)
= C_0\theta_0^2/(2\beta^2+\theta_0^2)$, the variance of sky
fluctuations described by a GACF after convolution with a single
gaussian beam.  The most likely value, and the 95\% and 99.7\% limits,
for ${\Delta}T_{\rm rms} = [\overline{C}(0)]^{1/2}$ are shown in
Figure~\ref{f11n}.  At small values of $\theta_0$, our experiment is
unable to distinguish between GACFs with different coherence angles
and so $\Delta T_{\rm rms}$ tends towards a constant value as
$\theta_0$ tends to zero.  At large values of $\theta_0$, the SuZIE
differencing scheme decreases the sensitivity to ${\Delta}T_{\rm rms}$
and so the limits rise steeply with increasing $\theta_0$.  Selecting
the value of $\theta_0=1\farcm1$ (the GACF to which this experiment is
most sensitive) yields upper limits on ${\Delta}T_{\rm rms}$ of:
\begin{equation}
\Delta T_{\rm rms}/T \le \left\{  \begin{array}{ll}
   1.6\times 10^{-5} \mbox{\hspace*{5mm} (95\% confidence)}\\
   1.9\times 10^{-5} \mbox{\hspace*{5mm} (99.7\% confidence)}
                        \end{array}
               \right.
\end{equation}
As $\theta_0$ tends to zero, the limits become:
\begin{equation}
\Delta T_{\rm rms}/T \le \left\{  \begin{array}{ll}
   1.2\times 10^{-5} \mbox{\hspace*{5mm} (95\% confidence)}\\
   1.5\times 10^{-5} \mbox{\hspace*{5mm} (99.7\% confidence)}
                        \end{array}
               \right.
\end{equation}

\placefigure{f11n}

\section{Discussion}

These results are compared with other limits on similar angular scales
and are used to set limits on sources of secondary anisotropy.

\subsection{Comparison with other measurements at arcminute scales}

Several other experiments have explored CMB anisotropies on similar
angular scales to the SuZIE measurements.  In general, the most
sensitive limits prior to this work have been obtained from
measurements at centimeter wavelengths.  These include: (i) the NCP
(North Celestial Pole) and RING measurements made with the Owens
Valley Radio Observatory (OVRO) 40\,m dish at a frequency of 20\,GHz
(\cite{r1}, \cite{m1}); (ii) a measurement made with the most compact
configuration of the Very Large Array at 15\,GHz (\cite{fo1}); (iii) a
limit set with the Australia Telescope at 8.7\,GHz (\cite{s1}).  At
millimeter wavelengths (90\,GHz), an upper limit to arcminute scale
anisotropy has been obtained by the White Dish experiment (\cite{tu1}).

Of these measurements, only the Owens Valley RING measurement has
detected structure.  Myers et al.\ (1993) report a detection of
structure with $2.3 < \Delta T_{\rm rms}/T < 4.5\times 10^{-5}$ (95\%
confidence limits) that is inconsistent with the 95\% upper limit of
$1.7\times10^{-5}$ obtained with the same instrument in the NCP
experiment (the RING measurement contains 96 108$''$ FWHM fields
compared to the 8 $108''$ fields observed in the NCP experiment).
Confusion from radio point sources is postulated as the cause of this
disagreement.  

\begin{figure}
\plotone{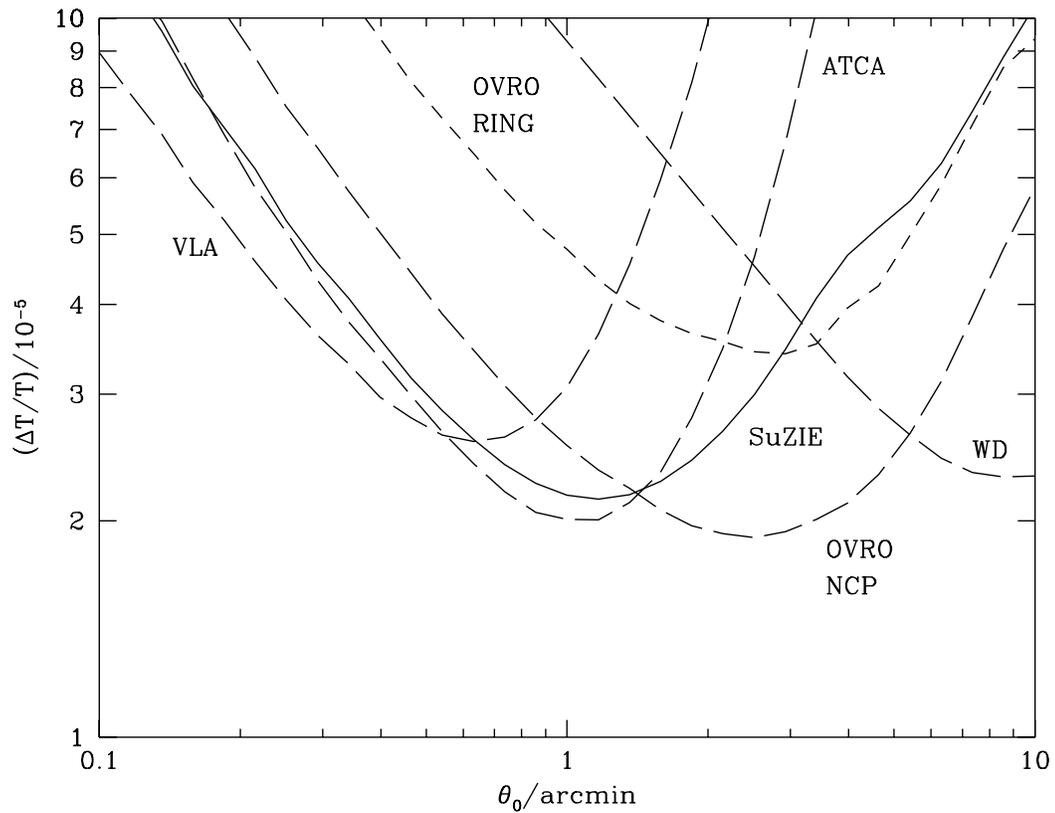}
\caption[]
{SuZIE 95\% confidence limits (solid line) to a gaussian
  autocorrelation function for $\Delta T/T$ anisotropies, as a
  function of $\theta_0$.  The corresponding 95\% confidence limits
  (long-dashed lines) from the Owens Valley Radio Observatory (OVRO)
  NCP measurement, the Very Large Array (VLA), the Australia Telescope
  Compact Array (ATCA) and the White Dish (WD) experiment are also
  shown (for references, see text).  The position of the likelihood
  peak obtained from the OVRO RING experiment is also shown
  (short-dashed line). }
\label{f12n} 
\end{figure}

The 95\% upper limits to a GACF obtained from each experiment, and the
SuZIE 95\% upper limits, are plotted as a function of $\theta_0$ in
Figure~\ref{f12n}.  At millimeter wavelengths, the SuZIE data sets the
most sensitive upper limits to GACFs with $\theta_0 \le 3'$.  The
SuZIE 95\% upper limit also lies below the RING detection for
coherence angles less than $2\farcm5$, consistent with the hypothesis
that the RING detection is due to a population of radio point sources.

\placefigure{f12n}

\subsection{Limits on populations of Sunyaev-Zel'dovich sources}

Inverse-Compton scattering of CMB photons by the hot gas in galaxy
clusters causes the Sunyaev-Zel'dovich effect (\cite{s2}, 1980).  The
spectral distortions that comprise the S-Z effect can be
conveniently sub-divided into a component arising from the thermal
motions of the electrons in the hot gas, and a component arising from
the kinematic velocity of the cluster with respect to the CMB rest
frame (\cite{r2}).

The non-relativistic form of the thermal effect can be written as a
frequency-dependent $\Delta T/T_{\rm CMB}$ distortion:
\begin{equation}
  \frac{\Delta T}{T_{\rm CMB}} = -\left[ x\coth\frac{x}{2} -4\right]y
  \label{e8.2.1}
\end{equation}
where $x=h\nu/kT_{\rm CMB}$ and the Compton $y$-parameter is given by
$y\sim \tau kT_e/m_ec^2$ for a cluster with optical depth $\tau$ to
Compton scattering, and average electron temperature, $T_e$.  The
thermal S-Z effect thus appears as a decrement in the CMB temperature
at wavelengths longer than $\sim 1.4$\,mm and an increment at shorter
wavelengths.  Clusters with strong X-ray emission typically have
values of $\tau=0.01$--0.02 and $T_e=8$--15\,keV.  In the
non-relativistic limit the cross-over occurs at $\lambda = 1.38$\,mm,
otherwise the exact point of zero-crossing depends on well-defined
relativistic corrections (\cite{r3}).

The kinematic effect has a spectral signature identical to that of
primordial CMB fluctuations and is given by:
\begin{equation}
  \frac{\Delta T}{T_{\rm CMB}} = \frac{v}{c} \tau
\end{equation}
where $v$ is the peculiar velocity of the cluster with respect to the
CMB rest frame.  In Cold Dark Matter models with density parameter
$\Omega=0.3$, the rms value of peculiar velocities is predicted to be
400\,km\,s$^{-1}$ with 10\% of clusters having $v_{\rm pec} \ge
700$\,km\,s$^{-1}$; for $\Omega=1$, the rms velocity is
700\,km\,s$^{-1}$ with a high velocity tail out to 2000\,km\,s$^{-1}$
(\cite{b9}).  Therefore, for a bright cluster with $\tau=0.02$ and
$T_e=10$\,keV, the kinematic effect is expected to be at least an
order of magnitude smaller than the thermal effect, except near
$\lambda=1.4$\,mm where the thermal effect is zero.

An important feature of both S-Z components is that the spectral
shapes and intensities are independent of redshift for clusters with
identical properties.  Thus theoretical predictions of number counts
of S-Z sources depend only on models of large-scale structure and the
evolution of cluster gas properties with redshift.

Several authors consider the contribution to $\Delta T/T$ anisotropy
from a population of X-ray clusters.  Bond (1995) and Markevitch et
al.\ (1992, 1994) have simulated maps of several square degrees of
sky, showing the expected contribution from the S-Z effect.  De Luca,
D\'{e}sert and Puget (1995), Markevitch et al.\ (1994), Bartlett and
Silk (1994) and Barbosa et al.\ (1996) have calculated the expected
number counts of S-Z sources as a function of flux and have explored
the dependence of this statistic on the assumed cosmological model.
Bond (1995) and De Luca et al.\ (1995) consider the contribution from
both the thermal and kinematic effects; Markevitch et al.\ (1992,
1994), Bartlett and Silk (1994) and Barbosa et al.\ (1996) consider
the thermal effect only.

The correlation function of cluster-induced anisotropies is strongly
dependent on the assumed cosmological model (see Figure~7 of \cite{b4}
for theoretical models of the cluster-induced $C_l$ power spectrum).
Experimental results such as those obtained with SuZIE are best
compared with theoretical models of such anisotropies via detailed
simulations of model skys, convolved with the appropriate beam
response.  However, since there is no significant detection of
structure in the SuZIE data, a simplistic comparison of the $\Delta
T/T$ limits from the GACF analysis to the models serves as a useful
guide to how closely current experimental limits are approaching
theoretical predictions.

Since the SuZIE beam size is well matched to the size of a typical
cluster at $z>0.1$, we use the 99.7\% confidence limit of $\Delta T/T
\le 2.6\times 10^{-5}$ (from Equation~\ref{e31}) obtained using a GACF with
a coherence angle of $1\farcm1$.  Converting to a value for the
$y$-parameter yields $\Delta y \le 2.5\times 10^{-5}$ at the 99.7\%
confidence level.  Ignoring the effects of the SuZIE differencing
scheme yields a limit to the S-Z flux at 142\,GHz from any cluster
that may be in the SuZIE fields of $|\Delta I_{\rm S-Z}| \le
7.3$\,mJy.  Scaling theoretical S-Z number counts at 400\,GHz
(\cite{b11}) to 142\,GHz yields a prediction of 1.2 clusters with
absolute flux greater than $ 7.3$\,mJy in the 0.06 square degrees
observed by SuZIE if $\Omega=0.3$.  If $\Omega=1$, 0.2 clusters per
0.06 square degrees are expected.  Thus in a low-$\Omega$ universe,
the SuZIE limits are comparable to the theoretical predictions.
Because the expected number of clusters is small, observations of
larger sky patches to a similar sensitivity level are required to
allow meaningful comparisons with models of the S-Z cluster
background. The results reported here indicate that such a test is
within reach in the near future.

\section{Conclusions}

Based on observations of two patches of sky covering a total area of
0.06 sq. deg. at 142\,GHz with $1\farcm7$ resolution, we have set 95\%
confidence limits of $\Delta T/T \le 2.1 \times 10^{-5}$ for CMB
anisotropies distributed with a gaussian autocorrelation function with
a $1\farcm1$ coherence angle.  These limits do not include the effects
of atmospheric correlations in the data, but a partial re-analysis
that accounts for these correlations has been shown to reduce the
upper limits.  Consequently the numbers we present here are
conservative.  These limits are comparable to the best limits obtained
from centimeter-wave observations on similar angular scales. Because
the SZ effect is brighter relative to CMB anisotropy at 142~GHz than
at centimeter wavelengths, these upper limits on CMB anisotropy
provide the most sensitive probe of the background of SZ fluctuations
to date.

The upper limits that we have obtained are comparable to fluctuations
from the Sunyaev-Zel'dovich effect in galaxy clusters predicted for a
low-$\Omega$ universe.  Observations of larger areas of sky at a
similar level of sensitivity, coupled to detailed simulations of model
skies that include the effects of our scan strategy, will
significantly constrain models of the formation of galaxy clusters.

%The sensitivity that we have achieved with this single-frequency
%instrument is limited by the presence of atmospheric noise in the data
%that is not adequately removed by differencing pixels in the focal
%plane.  Additionally, single-frequency observations are unable to
%unambiguously identify the sources of true anisotropy in the data.
%This is particularly important for measurements at these angular
%scales as several sources of anisotropy with different spectral
%signatures, such as primordial anisotropies and the S-Z effect, are
%expected to contribute comparable amounts of signal at these
%wavelengths.

An upgraded SuZIE instrument has now been constructed and is currently
being commissioned at the CSO.  This multi-frequency instrument makes
simultaneous measurements at 142, 217 and 268\,GHz and thus allows
atmospheric noise, which limits the sensitivity of the current system,
to be subtracted via correlation between different frequency channels.
The multi-frequency capability will also allow separation of primary
CMB anisotropies from secondary fluctuations such as those caused by
S-Z clusters.  This instrument will achieve sensitivity levels
significantly better than the results reported here, enabling us to
detect both primary and secondary anisotropies if they exist at levels
predicted by current theories.

\acknowledgments This work has been made possible by a grant from the
David and Lucile Packard Foundation, and by a National Science
Foundation grant \#AST-95-03226.  We thank Anthony Schinckel and the
entire staff of the CSO for their excellent support during the
observations.  The CSO is operated by the California Institute of
Technology under funding from the National Science Foundation,
Contract \#AST-93-13929.

\appendix
\section{Derivation of the residual atmospheric correlation function}
\label{a1}
If the data in the $i$th pixel of the $j$th scan is given by
$y_{i,j}=t_i + n_{i,j}$, where $t_i$ is the astronomical signal and
$n_{i,j}$ is the noise term, then the co-added signal as given by
Equations~\ref{e1} and~\ref{e2} can be written as:
\begin{equation}
  y_i = t_i + n_i
\end{equation}
where:
\begin{equation}
  n_i = \frac{\sum\limits_{j}^{} n_{i,j}/\sigma^2_{i,j}}{\sum\limits_{j}^{}
    1/\sigma^2_{i,j}}
\end{equation}
We ignore for the moment the simplification that can be applied to
this particular data set, that $\sigma_{i,j}=\sigma_j$ and is
identical for all points within a single scan.  

The correlation function of the co-added data, $S_r=\langle y_i
y_{i+r}\rangle$, is then:
\begin{eqnarray}
  S_r & = & \langle t_i t_{i+r} \rangle + \langle n_i n_{i+r} \rangle \\
      & = & {\cal C}_r + {\cal N}_r
\end{eqnarray}
where ${\cal C}_r$ is the correlation function of the true
astronomical signal and ${\cal N}_r$ is the residual contribution to
the correlation function from noise correlated within a scan.
Substituting the full expression for $n_i$ into the definition of the
correlation function given in Equation~\ref{e20} yields:
\begin{equation}
  {\cal N}_r = \frac{\sum\limits_{i=1}^{N-r} 
    \left( \sum\limits_{j}^{} n_{i,j}/\sigma^2_{i,j} \right)
      \left( \sum\limits_{j'}^{} n_{i+r,j'}/\sigma^2_{i+r,j'} \right) }
      {\sum\limits_{i=1}^{N-r} 
        \left( \sum\limits_{j}^{} 1/\sigma^2_{i,j} \right)
          \left(\sum\limits_{j}^{} 1/\sigma^2_{i+r,j'}\right)}
\end{equation}

Since we assume that $n_{i,j}$ is uncorrelated between scans, the
numerator of this expression is zero except when $j=j'$.  Thus:
\begin{equation}
  {\cal N}_r = \frac{\sum\limits_{i=1}^{N-r} \sum\limits_{j}^{} 
    \left( n_{i,j}/\sigma^2_{i,j} \right)
      \left( n_{i+r,j}/\sigma^2_{i+r,j} \right)}
      {\sum\limits_{i=1}^{N-r} 
        \left( \sum\limits_{j}^{} 1/\sigma^2_{i,j} \right)
          \left( \sum\limits_{j'}^{} 1/\sigma^2_{i+r,j'} \right) }
\end{equation}
Reversing the order of summation in the numerator yields:
\begin{equation}
  {\cal N}_r = \frac{ \sum\limits_{j}^{} {\cal N}_{r,j} \times 
    \sum\limits_{i=1}^{N-r} (1/\sigma^2_{i,j}) (1/\sigma^2_{i+r,j})}
  {\sum\limits_{i=1}^{N-r} \left( \sum\limits_{j}^{} 1/\sigma^2_{i,j} \right) 
    \left( \sum\limits_{j'}^{} 1/\sigma^2_{i+r,j'} \right)}
    \label{eA7}
\end{equation}
where ${\cal N}_{r,j}$ is the correlation function of the noise in the
$j$th scan and is given by:
\begin{equation}
  {\cal N}_{r,j} = \frac{\sum\limits_{i=1}^{N-r}
    (n_{i,j}/\sigma^2_{i,j})
    (n_{i+r,j}/\sigma^2_{i+r,j})}{\sum\limits_{i=1}^{N-r}
    (1/\sigma^2_{i,j}) (1/\sigma^2_{i+r,j})}
\end{equation}
Since in this data set $\sigma_{i,j}$ is constant for all values of
$i$, Equation~\ref{eA7} simplifies to:
\begin{equation}
  {\cal N}_r = 
  \frac{\sum\limits_{j}^{} {\cal N}_{r,j}/\sigma^4_j}{(\sum\limits_{j}^{} 
    1/\sigma^2_j)^2}
\end{equation}

\clearpage

% Now comes the reference list.  In this document, we used \cite to call
% out citations, so we must use \bibitem in the reference list, which
% means we use the LaTeX thebibliography environment.  Please note that
% \begin{thebibliography} is followed by a null argument.  If you forget
% this, mayhem ensues, and LaTeX will say "Perhaps a missing item?" when
% you run it.  Do not call us, do not send mail when this happens.  Put
% the silly {} after the \begin{thebibliography}.
%
% Each reference has a \bibitem command to define the citation format
% to be placed in the text (in []) and the symbolic tag used for 
% cross referencing (in {}).
%
% See sample1.tex, or the AASTeX guide, for an alternative to the \cite-
% \bibitem command.

\clearpage

% Tables here
 
\begin{deluxetable}{ccccc}
  \footnotesize
  \tablecaption{ Locations of the two fields observed in April 1994\label{t1}}
  \tablewidth{0pt}
  \tablehead{  
    & \multicolumn{2}{c}{Equatorial} 
    & \multicolumn{2}{c}{Galactic}   \\
    \cline{2-5}
    \colhead{} & \colhead{RA (1950)} & \colhead{Dec (1950)} &
    \colhead{l} & \colhead{b}
    } 
  \startdata
  Field 1 & $10^{\rm h} 21^{\rm m} 49^{\rm s}$ & $4^\circ 04' 23''$
          & $240\fdg01$                        & $47\fdg89$ \nl
  Field 2 & $16^{\rm h} 30^{\rm m} 17^{\rm s}$ & $5^\circ 56' 00''$
          & $21\fdg35$                         & $33\fdg37$ \nl
  \enddata
\end{deluxetable}

%\begin{deluxetable}{ccccc}
%  \footnotesize \tablecaption{Magnitudes in $\mu$K of (i) the linear
%    drift removed from each channel and (ii) the correlation of the
%    data from each chop to the temperature of the 300\,mK stage.  The
%    corresponding maximum temperature excursion, $\Delta T_{\rm max}$
%    is also shown. \label{t2}}
%  \tablewidth{0pt}
%  \tablehead{
%    & Chop & Drift & Temperature correlation & $\Delta T_{\rm max}$ \\
%    & & \colhead{[$\mu$K\,s$^{-1}$]} &
%    \colhead{[nK$_{\rm CMB}/nK$_{\rm 300\,mK}$]} & \colhead{[$\mu$K$_{\rm CMB}$]}
%    } 
%  \startdata
%  Field 1 & $t_{123}$ &  24 & 23.8 & 1.9 \nl
%  & $ d_{31}$ &  16 & 11.3 & 0.9 \nl
%  & $t_{456}$ &  48 & 25.0 & 2.0 \nl
%  & $ d_{64}$ &  -3 & 30.0 & 2.4 \nl
%  &           &     &     \nl
%  Field 2 & $t_{123}$ &   8 & 22.5 & 1.8 \nl
%  & $ d_{31}$ & -39 & 7.5  & 0.6 \nl
%  & $t_{456}$ & -61 & 10.0 & 0.8 \nl
%  & $ d_{64}$ & -27 & 44.0 & 3.5 \nl
%  \enddata
%\end{deluxetable}

\begin{deluxetable}{cccc}
  \footnotesize \tablecaption{Average magnitudes in $\mu$K\,s$^{-1}$
 of the linear drift removed from each channel. \label{t2}}
  \tablewidth{0pt}
  \tablehead{
    & Chop & Mean drift & Rms drift \nl
%    &      & Drift & Drift \nl
    & & \colhead{[$\mu$K\,s$^{-1}$]} & \colhead{[$\mu$K\,s$^{-1}$]} \nl
    } 
  \startdata
  Field 1   & $t_{123}$ &  25 & 118 \nl
(190 scans) & $ d_{31}$ &  16 & 45 \nl
            & $t_{456}$ &  49 & 197 \nl
            & $ d_{64}$ &  -6 & 41 \nl
            &           &     \nl
  Field 2   & $t_{123}$ &   7 & 56 \nl
(277 scans) & $ d_{31}$ & -39 & 103 \nl
            & $t_{456}$ & -63 & 240 \nl
            & $ d_{64}$ & -28 & 113 \nl
  \enddata
\end{deluxetable}

\begin{deluxetable}{cccc}
  \footnotesize \tablecaption{Calculated value of $\chi^2$ for each
    difference and each field.  The probability of the $\chi^2$ value
    being exceeded is also shown.
    \label{t2b}}
  \tablewidth{0pt}
  \tablehead{
    & \colhead{difference} & \colhead{$\chi^2$ (46 d.o.f.)} 
                           & \colhead{$P(>\chi^2)$} 
    } 
  \startdata
  Field 1 & $t_{123}$ & 55.6  & 0.157 \nl
  & $ d_{31}$ & 29.1  & 0.976 \nl
  & $t_{456}$ & 31.7  & 0.947 \nl
  & $ d_{64}$ & 40.4  & 0.705 \nl
  &           &       &       \nl
  Field 2 & $t_{123}$ & 39.3  & 0.746 \nl
  & $ d_{31}$ & 35.9  & 0.857 \nl
  & $t_{456}$ & 45.5  & 0.491 \nl
  & $ d_{64}$ & 33.5  & 0.915 \nl
  \enddata
\end{deluxetable}

\begin{deluxetable}{ccccc}
  \footnotesize \tablecaption{Results of maximum likelihood analysis to
    determine the magnitude of excess variance in each data set.
    \label{t3}}
  \tablewidth{0pt}
  \tablehead{
    & & & \multicolumn{2}{c}{Upper confidence limits, $\sigma_u$ [$\mu$K]} \\
    \cline{4-5}
    \colhead{Field} & \colhead{Chop} & \colhead{Peak [$\mu$K]} 
                    & \colhead{95\%} & \colhead{99.7\%}
    } 
  \startdata
  Field 1 & $t_{123}$ & 47 & 98  & 132 \nl
  & $ d_{31}$ & 0  & 79  & 120 \nl
  & $t_{456}$ & 1  & 98  & 148 \nl
  & $ d_{64}$ & 0  & 109 & 156 \nl
  &           &    &     &     \nl
  Field 2 & $t_{123}$ & 0  & 52  & 77  \nl
  & $ d_{31}$ & 0  & 95  & 140 \nl
  & $t_{456}$ & 27 & 116 & 162 \nl
  & $ d_{64}$ & 1  & 94  & 140 \nl
  \enddata
  
\end{deluxetable}

\clearpage

% And finally, we must deal with the figures.

%% New Figure 1

%% New Figure 2

%% Figure 3

%% Figure 4

%% Figure 5

%% Figure 6

%% Figure 7

%% Figure 8

%% Figure 9

%% Figure 10

%% Figure 11

%% Figure 12

% That's all, folks.


\begin{thebibliography}{}
\bibitem[Bahcall, Cen and Gramann 1994]{b9} Bahcall, N. A., Cen,
R., and Gramann, M. 1994, \apjl, 430, L13
\bibitem[Barbosa et al.\ 1996]{b11} Barbosa,~D., Bartlett,~J. G.,
Blanchard,~A., and Oukbir,~J., 1996, \aap, in press
\bibitem[Bartlett and Silk 1994]{b1} Bartlett, J. G., and
Silk, J. 1994, \apj, 423, 12
\bibitem[Becker, White and Edwards 1991]{b2} Becker, R. H., White, R. L.,
and Edwards, A. L. 1991, \apjs, 75, 1
\bibitem[Berger 1985]{b10} Berger, J., 1985, Statistical Decision Theory and
Bayesian Analysis (New York: Springer-Verlag)
\bibitem[Bond et al.\ 1991a]{b3} Bond, J. R., Efstathiou, G., Lubin,
P. M., and Meinhold, P. R. 1991, \prl, 66, 2179
\bibitem[Bond 1995]{b4} Bond, J. R. 1995, in Cosmology and Large Scale
Structure, ed. R. Schaeffer, Elsevier Science Publishers, Netherlands
\bibitem[Bunn and Sugiyama 1995]{b8} Bunn, E. F., and Sugiyama, N. 1995,
\apj, 446, 49
\bibitem[De Luca, D\'{e}sert and Puget 1995]{d1} De Luca, A.,
D\'{e}sert, F. X., and Puget, J. L. 1995, \aap, 300, 335
\bibitem[Fomalont et al.\ 1993]{fo1} Fomalont, E. B., Partridge, R. B.,
Lowenthal, J. D., and Windhorst, R. A. 1993, \apj, 404, 8
\bibitem[Ganga et al.\ 1997a]{g1} Ganga, K. M., Ratra, B., Church,
S. E., Sugiyama, N., Holzapfel, W. L., Mauskopf, P. D., Wilbanks,
T. M., Ade, P. A. R., and Lange, A. E. 1997a, \apj, submitted
\bibitem[Ganga et al.\ 1997b]{g4} Ganga, K. M., Ratra, B.,
Gunderson,~J.~O., and Sugiyama, N., 1997b, \apj, submitted
\bibitem[G\'{o}rski et al.\ 1995]{g3} G\'{o}rski, K. M., Ratra, B.,
Sugiyama, N., and Banday, A. J. 1995, \apjl, 444, L65
\bibitem[Glezer, Lange \& Wilbanks~1992]{glezer1} 
  Glezer, E.N., Lange, A.E., \& Wilbanks, T.M.~1992, Applied Optics, 31, 7214
\bibitem[Griffin and Orton 1993]{g2} Griffin, M. J. and Orton,
G. S. 1993, Icarus, 105, 537
\bibitem[Holzapfel et al.\ 1997a]{h1} Holzapfel, W. L., Arnaud, M.,
Ade, P. A. R., Church, S. E., Fischer, M. L., Mauskopf, P. D.,
Rephaeli, Y., Wilbanks, T. M., and Lange, A. E. 1997a, \apj, in press
\bibitem[Holzapfel et al.\ 1997b]{w4} Holzapfel, W. L., Wilbanks,
T. M., Ade., P. A. R., Church, S. E., Fischer, M. L., Mauskopf, P. D.,
Osgood, D. E., and Lange, A. E. 1997b, \apj, in press
\bibitem[Jones and Forman 1984]{j1} Jones, C., and Forman, W., 1984,
\apj, 276, 38
\bibitem [Lawrence et al.\ 1988]{la1} Lawrence, C.R.,
Readhead,~A.C.S., and Myers,~S.T., 1988, in The Post-Recombination
Universe, eds. N. Kaiser and A.N.~Lasenby, NATO ASI Series, Kluwer
Academic Publishers, Dordrecht 
\bibitem[Mather et al.\ 1994]{m3} Mather, J. C.  et al.\ 1994, \apj,
420, 439
\bibitem[Markevitch et al.\ 1992]{m2} Markevitch, M., Blumenthal, G. R.,
Forman, W., and Jones, C. 1992, \apj, 395, 326
\bibitem[Markevitch et al.\ 1994]{m4} Markevitch, M., Blumenthal, G. R.,
Forman, W., Jones, C., Sunyaev, R. A. 1994, \apj, 426, 1
\bibitem[Myers et al.\ 1993]{m1} Myers, S. T., Readhead, A. C. S., and
Lawrence, C. R. 1993, \apj, 405, 8
\bibitem[Ratra et al. 1997]{r4} Ratra, B., Sugiyama, N., Banday,
A. J. and G\'{o}rski, K. M. 1997, Princeton preprint PUPT-1558+1559,
\apj, 481, in press
\bibitem[Readhead et al.\ 1989]{r1} Readhead, A. C. S., Lawrence,
C. R., Myers, S. T., Sargent, W. L. W., Hardebeck, H. E., and Moffett,
A. T. 1989, \apj, 346, 566
\bibitem[Rephaeli and Lahav 1991]{r2} Rephaeli, Y. and Lahav, O. 1991,
\apj, 372, 21
\bibitem[Rephaeli 1995]{r3} Rephaeli, Y. 1995, \apj, 445, 33
\bibitem[Subramanyan et al.\ 1993]{s1} Subrahmanyan, R., Ekers, R. D.,
Sinclair, M., and Silk, J. 1993, \mnras, 263, 416 
\bibitem[Sunyaev and Zel'dovich 1972]{s2} Sunyaev, R.A., and
Zel'dovich, Ya. B., 1972, Comments Astrophys. Space Phys., 4, 173
\bibitem[Sunyaev and Zel'dovich 1980]{s3} Sunyaev, R.A., and
Zel'dovich, Ya. B., 1980, \mnras, 190, 413
\bibitem[Tucker et al.\ 1993]{tu1} Tucker, G. S., Griffin, G. S.,
Nguyen, H. T., and Peterson, J. B. 1993, \apjl, 419, L45
\bibitem[White et al.\ 1994]{w1} White, M., Scott, D., and Silk,
J. 1994, \araa, 32, 319
\bibitem[Wilbanks et al.\ 1990]{w2} Wilbanks, T. M., Devlin, M.,
Lange, A. E., Sato, S., Beeman, J. W., and Haller, E. E. 1990, IEEE
Trans. Nucl. Sci., 37, 566
\bibitem[Wilbanks et al.\ 1994]{w3} Wilbanks, T. M., Ade., P. A. R.,
Fischer, M. L., Holzapfel, W. L., and Lange, A. E. 1994, \apjl, 427,
L75
\end{thebibliography}
\end{document}